\documentclass[10pt,a4paper]{article}

\usepackage[margin=2.45cm]{geometry}
\usepackage{amsmath,amssymb,mathtools}
\usepackage{graphicx}
\usepackage{bm}
\usepackage{booktabs}
\usepackage{needspace}
\usepackage{placeins}
\usepackage{float}
\usepackage{slashed}
\usepackage{xcolor}
\usepackage[numbers,sort&compress]{natbib}
\usepackage[colorlinks=true,linkcolor=blue,citecolor=blue,urlcolor=blue]{hyperref}

\graphicspath{{figs/}}
\newcommand{\includecharmplot}[2][]{%
  \IfFileExists{#2}{%
    \includegraphics[#1]{#2}%
  }{%
    \fbox{\parbox[c][0.18\textheight][c]{0.88\linewidth}{%
      \centering Original charm-channel figure required:\\[2pt]
      \texttt{\detokenize{#2}}%
    }}%
  }%
}
\newcommand{\includebottomplot}[2][]{%
  \IfFileExists{#2}{%
    \includegraphics[#1]{#2}%
  }{%
    \fbox{\parbox[c][0.18\textheight][c]{0.88\linewidth}{%
      \centering Bottom-channel figure asset not available:\\[2pt]
      \texttt{\detokenize{#2}}%
    }}%
  }%
}
\newcommand{\GG}{\langle g_s^2 G^2\rangle}
\newcommand{\mixed}{\langle\bar s g_s\sigma\!\cdot\! G s\rangle}

\newcommand{\GeV}{\ \mathrm{GeV}}
\newcommand{\GeVsq}{\ \mathrm{GeV}^2}
\newcommand{\MeV}{\ \mathrm{MeV}}
\newcommand{\Pproj}{\mathcal{P}^{(2)}}

\newcommand{\qq}{\langle\bar s s\rangle}
\newcommand{\dd}{\mathrm d}
\newcommand{\Huv}{H_{uv}}

\begin{document}

\title{Spin-2 $QQ\bar Q\bar s$ Tetraquarks in QCD Sum Rules}

\author{%
Tarik Akan$^{1}$%
\thanks{%
\href{mailto:tarik.akan@bozok.edu.tr}
{tarik.akan@bozok.edu.tr} (Corresponding author)}
\\[2mm]
$^{1}$Physics Department, Yozgat Bozok University,
66100 Yozgat, Turkey
}

\date{}

\maketitle
\begin{abstract}
The $J^{P}=2^{+}$ $cc\bar c\bar s$ and $bb\bar b\bar s$ tetraquarks are studied in parallel within the two-point QCD Sum Rules framework using the same symmetric axial-vector$\otimes$axial-vector interpolating current. The current-specific OPE sectors retained through dimension seven within the stated background-field truncation are evaluated with a common symbolic construction, with the bottom channel obtained by the heavy-flavour replacement $c\to b$. The same pole-dominance and OPE-convergence criteria are imposed in both channels. The adopted charm region is $\sqrt{s_0}=5.7\GeV$ with $3.6\le M_B^2\le4.3\GeVsq$, whereas the bottom region is $\sqrt{s_0}=14.3\GeV$ with $9.8\le M_B^2\le10.8\GeVsq$. At representative Borel points, the extracted masses are $5.239\GeV$ and $13.832\GeV$, respectively. After the QCD-input, continuum-threshold, and Borel-window variations are combined, the final results are $m_{cc\bar c\bar s,\,2^+}=5.239\pm0.061\GeV$ and $m_{bb\bar b\bar s,\,2^+}=13.832\pm0.059\GeV$.
\end{abstract}

\section{Introduction}
\label{sec:intro}

QCD Sum Rules provide a direct connection between hadronic observables and quark-gluon degrees of freedom through the operator product expansion (OPE), in which perturbative Wilson coefficients are combined with nonperturbative vacuum condensates \cite{Shifman:1978bx,Shifman:1978by,Reinders:1984sr,Colangelo:2000dp}. Multiquark systems provide a useful setting for this framework because the extracted masses can be tested against pole dominance, OPE convergence, and Borel stability while retaining explicit sensitivity to the interpolating current. Experimental studies of hidden- and fully-heavy structures have further motivated theoretical investigations of compact tetraquarks and hadronic molecules \cite{Aaij:2020fnh,Chen:2016qju,Brambilla:2019esw}.

Triply heavy tetraquarks of the form $QQ\bar Q\bar s$ are especially suitable for a controlled heavy-flavour comparison. The $cc\bar c\bar s$ and $bb\bar b\bar s$ channels have the same light-flavour content and differ only through the heavy-quark species. Consequently, applying the same current, OPE truncation, numerical acceptance criteria, and uncertainty prescription to both systems provides a direct test of how the sum rule changes from charm to bottom.

For the $cc\bar c\bar s$ system, Ref. \cite{An:2026TriplyHeavy} reports an S-wave $J^{P}=2^{+}$ state at $5.489\GeV$ in a nonrelativistic quark model and identifies $J/\psi D_s^{*}$ as a dominant S-wave rearrangement channel. The same study predicts the corresponding $bb\bar b\bar s$, $J^{P}=2^{+}$ state at $15.238\GeV$ and identifies $\Upsilon B_s^{*}$ as its allowed S-wave rearrangement channel. QCD Sum Rules have also been applied to the $cc\bar c\bar s$ flavour content in the scalar sector. Ref. \cite{Xu:2025} reports masses near $4.94$-$5.08\GeV$ for several $J^P=0^+$ currents. These values provide phenomenological context, but no spin ordering is inferred from comparisons between different currents and quantum numbers.

Accordingly, the $J^{P}=2^{+}$ $cc\bar c\bar s$ and $bb\bar b\bar s$ states are analyzed on exactly the same footing. A single generic $QQ\bar Q\bar s$ current with $Q=c,b$ is used, and the current-specific sectors retained through $D=7$ are evaluated within the same background-field truncation. The numerical results are then presented in a parallel format: the six diagnostics used for the charm analysis are paired directly with the corresponding bottom-channel diagnostics. Section \ref{sec:method} defines the current, spin-2 projection, OPE organization, and sum-rule criteria. Section \ref{sec:results} presents the paired numerical analysis and the extracted masses. The Appendix collects the symbolic densities and related technical definitions in a heavy-flavour-generic notation.

\section{\texorpdfstring{QCD Sum Rules Methodology}{QCD Sum Rules Methodology}}
\label{sec:method}

In this work, the standard two-point QCD Sum Rules construction is used. The hadronic representation of the correlation function is matched to its OPE representation, followed by Borel transformation and continuum subtraction. The related parameters are fixed before the mass is quoted.

\subsection{Interpolating current}
\label{subsec:current-wick}

The interpolating field is the symmetric rank-two combination of an axial-vector diquark
and an axial-vector antidiquark,
\begin{equation}
  J_{\mu\nu}^{(Q)}=\frac{1}{\sqrt2}\,\epsilon^{abc}\epsilon^{ade}
  \Big[(Q_b^{T}C\gamma_\mu Q_c)(\bar Q_d\gamma_\nu C\bar s_e^{T})
       +(\mu\!\leftrightarrow\!\nu)\Big],\qquad Q=c,b,
  \label{eq:current}
\end{equation}
where $C$ is the charge-conjugation matrix and $a,\dots,e$ are colour indices. The same local operator is used for the charm and bottom channels; only the heavy flavour and its numerical input are changed. The superscript $(Q)$ is suppressed below whenever the same equation applies to both channels. The choice of $C\gamma_\mu$ is constrained by Fermi statistics. For anticommuting heavy fields $Q_b^{T}MQ_c=-Q_c^{T}M^{T}Q_b$, so a symmetric $C\Gamma$ (where $\Gamma$ denotes the Dirac matrix structure) gives a colour-antisymmetric diquark under contraction with $\epsilon^{abc}$. In particular,
\begin{equation}
  \epsilon^{abc}\,Q_b^{T}C\gamma_5 Q_c\equiv 0,
  \label{eq:scalarzero}
\end{equation}
whereas the axial-vector structure is allowed.

The two-point function and its spin-2 projection are
\begin{equation}
  \Pi_{\mu\nu,\alpha\beta}(p)=i\!\int\! d^4x\,e^{ipx}
  \langle0|T\{J_{\mu\nu}(x)J^{\dagger}_{\alpha\beta}(0)\}|0\rangle,
  \label{eq:correlator}
\end{equation}
\begin{equation}
  \Pi_2(p^2)=\tfrac15\,\Pproj{}^{\,\mu\nu,\alpha\beta}\Pi_{\mu\nu,\alpha\beta},
  \label{eq:projection}
\end{equation}
where the spin-2 projector is
\begin{equation}
  \Pproj_{\mu\nu,\alpha\beta}=\tfrac12\big(\tilde g_{\mu\alpha}\tilde g_{\nu\beta}
  +\tilde g_{\mu\beta}\tilde g_{\nu\alpha}\big)
  -\tfrac13\tilde g_{\mu\nu}\tilde g_{\alpha\beta},\qquad
  \tilde g_{\mu\nu}=-g_{\mu\nu}+\frac{p_\mu p_\nu}{p^2}.
  \label{eq:projector}
\end{equation}
The projector is symmetric, transverse, traceless, and idempotent.
The complete contraction of the projector with itself is written as
\begin{equation}
 \Pproj_{\mu\nu,\alpha\beta}\,
 \mathcal{P}^{(2)\,\mu\nu,\alpha\beta}
 =5.
 \label{eq:projector-contraction}
\end{equation}
where the colon denotes complete contraction over both Lorentz-index pairs. Here, the value $5$ is the trace of the spin-2 projector and equals the number $2J+1=5$ of physical polarization states of a massive $J=2$ particle. Consequently, the factor $1/5$ in Eq. \eqref{eq:projection} normalizes the projector contraction so that the spin-2 invariant amplitude $\Pi_2(p^2)$ is isolated.

The coupling to the lowest tensor state is defined by
\begin{equation}
 \langle0|J_{\mu\nu}|T(p)\rangle=\lambda_{2^+}\,\varepsilon_{\mu\nu}(p),
 \label{eq:couplingdef}
\end{equation}
where $\lambda_{2^+}$ is the pole residue of the lowest tensor state and $\varepsilon_{\mu\nu}(p)$ is its spin-2 polarization tensor. The corresponding pole contribution to the projected invariant amplitude is
\begin{equation}
 \Pi_2^{\rm had}(p^2)=\frac{\lambda_{2^+}^{2}}{m_{2^+}^{2}-p^2}+\cdots, 
 \label{eq:hadronicpole}
\end{equation}
where $\cdots$ denotes higher resonances and continuum states.

Light- and heavy-quark propagators are expanded in the background field to generate the retained current-specific OPE sectors
\begin{equation}
  \rho_2^{\rm OPE}(s)=\rho_0^{(2)}(s)+\rho_3^{(2)}(s)+\rho_4^{(2)}(s)
  +\rho_5^{(2)}(s)+\rho_6^{(2)}(s)+\rho_7^{(2)}(s),
  \label{eq:rhoope-decomp}
\end{equation}
where $\rho_D^{(2)}$ denotes the contribution to the spin-2 projected spectral density, associated with OPE dimension $D$. Here, $D=0$ is the perturbative term, while $D=3,4,5,6,7$ denote the $\langle\bar s s\rangle$, $\langle g_s^2G^2\rangle$, $\langle\bar s g_s\sigma\!\cdot\!Gs\rangle$, factorized four-quark, and $\langle\bar s s\rangle\langle g_s^2G^2\rangle$ sectors, respectively. 
%However, the independent triple-gluon condensate $\langle g_s^3G^3\rangle$ is outside the adopted background-field truncation and is not assumed to vanish.  
%Because the perturbative contribution fixes the normalization of every subsequent Borel moment, its 
As an example, $D=0$ construction is displayed explicitly below. The full numerator polynomial and the higher-dimensional densities are given in Appendix \ref{app:full-symbolic}.

At $D=0$ all four quark lines are replaced by their free propagators. For either heavy flavour $Q=c,b$,
\begin{equation}
  S_Q^{(0)}(k)=\frac{\gamma\!\cdot\! k+m_Q}{k^2-m_Q^2+i0},\qquad
  S_s^{(0)}(k)=\frac{\gamma\!\cdot\! k+m_s}{k^2+i0}+\mathcal O(m_s^2),\qquad
  \widetilde S_q\equiv C S_q^T C .
  \label{eq:d0prop}
\end{equation}
Here, $S_Q^{(0)}$ and $S_s^{(0)}$ denote the free heavy- and strange-quark propagators. The three heavy propagators carry momenta $k_1,k_2,k_3$, and $k_4=p-k_1-k_2-k_3$ is assigned to the strange propagator. The strange propagator is expanded through first order in $m_s$. Using the central masses in Table \ref{tab:inputs} as power-counting indicators,
\begin{equation}
 \frac{m_s}{m_c}\simeq0.073,\qquad
 \frac{m_s}{m_b}\simeq0.022,\qquad
 \left(\frac{m_s}{m_c}\right)^2\simeq5.4\times10^{-3},\qquad
 \left(\frac{m_s}{m_b}\right)^2\simeq5.0\times10^{-4}.
 \label{eq:msmQ-hierarchy}
\end{equation}
Thus, the same linear-$m_s$ truncation is parametrically at least as well controlled in the bottom channel as in the charm channel. These ratios are used only for power counting because the quark masses are quoted at their conventional reference scales.

The color structure of each current is conveniently written as
\begin{equation}
A_{bc;de}\equiv\sum_a\epsilon^{abc}\epsilon^{ade}
=\delta_{bd}\delta_{ce}-\delta_{be}\delta_{cd},
\label{eq:colorA}
\end{equation}
where $A_{bc;de}$ denotes the antisymmetric color tensor generated by the two Levi-Civita symbols. Because the two heavy-quark fields in the diquark are identical, Wick's theorem produces two inequivalent contractions. In the \emph{direct} contraction, the heavy-quark color indices are paired without interchange, $b\to b'$ and $c\to c'$. Its pure color factor is
\begin{equation}
C_{\rm dir}
=\sum_{b,c,d,e}A_{bc;de}A_{bc;de}=12.
\label{eq:Cdir}
\end{equation}

By contrast, in the \emph{exchange} contraction, the two identical heavy-quark lines are interchanged, $b\to c'$ and $c\to b'$. Since $A_{cb;de}=-A_{bc;de}$, the corresponding color contraction is
\begin{equation}
C_{\rm ex}
=\sum_{b,c,d,e}A_{bc;de}A_{cb;de}=-12.
\label{eq:Cex}
\end{equation}
The minus sign in $C_{\rm ex}$ is therefore a consequence of the antisymmetric color tensor and is not the fermionic Wick sign. Wick's theorem supplies a second, independent minus sign for the exchanged contraction of two identical fermion fields. Consequently, the color factor multiplying the complete Wick-contracted correlator is
\begin{equation}
C_{\rm Wick}
=C_{\rm dir}-C_{\rm ex}=12-(-12)=24,
\label{eq:Cwick}
\end{equation}
where $C_{\rm dir}$ denotes the no-interchange color contraction, $C_{\rm ex}$ denotes the color factor after exchanging the identical heavy-quark lines, and $C_{\rm Wick}$ is the net factor after the relative fermionic sign required by Wick's theorem has been included.

For later use, let
\begin{equation}
 \mathcal W[X_1,X_2,X_3;Y]
 \label{eq:Wdef}
\end{equation}
denote the complete spin-2 projected direct-minus-exchange Wick contraction generated by Eq. \eqref{eq:current}, with the three heavy-quark propagators replaced by $X_1,X_2,X_3$ and the strange propagator by $Y$. The perturbative sector is therefore
\begin{equation}
 \Pi_0^{(2)}(p^2)=\mathcal W\!\left[S_Q^{(0)},S_Q^{(0)},S_Q^{(0)};S_s^{(0)}\right],
 \qquad
 \rho_0^{(2)}(s)=\frac{1}{\pi}\operatorname{Im}\Pi_0^{(2)}(s+i0).
 \label{eq:d0-wick-origin}
\end{equation}

The symmetrized tensor current contributes a factor two, while the definition of the spin-2 invariant contributes $1/5$. Therefore the free projected kernel carries the current-specific prefactor
$C_{\rm Wick}\times 2\times(1/5)=48/5$:
\begin{equation}
\Pi^{(0)}_2(p^2)=\frac{48}{5}\,i\!\int\!\frac{d^4k_1\,d^4k_2\,d^4k_3}{(2\pi)^{12}}
\,\mathcal P^{(2)\,\mu\nu,\alpha\beta}
\operatorname{Tr}\!\left[\gamma_\mu S_Q^{(0)}(k_1)\gamma_\alpha\widetilde S_Q^{(0)}(k_2)\right]
\operatorname{Tr}\!\left[\gamma_\nu\widetilde S_s^{(0)}(k_4)\gamma_\beta S_Q^{(0)}(k_3)\right] .
\label{eq:d0trace}
\end{equation}
\subsection{\texorpdfstring{Explicit perturbative $D=0$ construction}{Explicit perturbative D=0 construction}}
\label{subsec:d0-method}

Writing the four denominators as
\begin{equation}
D_i=k_i^2-m_Q^2+i0\quad(i=1,2,3),\qquad D_4=k_4^2+i0,
\label{eq:d0den}
\end{equation}
where $D_{1,2,3}$ are the three heavy-quark denominators and $D_4$ is the strange-quark denominator in the linear-$m_s$ convention. Their product is combined with four Feynman parameters,
\begin{equation}
\frac{1}{D_1D_2D_3D_4}
=6\!\int_0^1\!\prod_{i=1}^4 dx_i\,
\delta\!\left(1-\sum_{i=1}^4x_i\right)
\frac{1}{\bigl(\sum_i x_iD_i\bigr)^4},
\label{eq:d0feynman}
\end{equation}
where the Feynman parameters satisfy $x_i\ge0$ and $\sum_{i=1}^4x_i=1$. Completing the three loop-momentum squares produces the four-line combinations
\begin{align}
U_4&=x_1x_2x_3+x_1x_2x_4+x_1x_3x_4+x_2x_3x_4,\\
P_4&=x_1x_2x_3x_4,\qquad
\bar m_4^{\,2}=m_Q^2\frac{(x_1+x_2+x_3)U_4}{P_4},
\label{eq:d0symanzik}
\end{align}
where $U_4$ and $P_4$ are the four-line polynomials and $\bar m_4^{\,2}$ is the associated effective mass parameter. For the projected tensor channel, the exact projective map is
\begin{equation}
 x_1=\frac{vw\,t_4}{W_4},\quad x_2=\frac{uw\,t_4}{W_4},\quad
 x_3=\frac{uv\,t_4}{W_4},\quad x_4=\frac{uvw}{W_4},\qquad
 \left|\frac{\partial(x_1,x_2,x_3)}{\partial(u,v,w)}\right|
 =\frac{u^2v^2w^2t_4^2}{W_4^4},
\label{eq:d0projective}
\end{equation}
where $t_4=1-u-v-w$ and $W_4=uvw+uv\,t_4+uw\,t_4+vw\,t_4$. The map gives $U_4=u^2v^2w^2t_4^2/W_4^3$ and $P_4=u^3v^3w^3t_4^3/W_4^4$. Accordingly, the four-line kinematic polynomial is defined as
\begin{equation}
 s-\bar m_4^{\,2}=\frac{K_{4}}{uvw},\qquad
 K_{4}(u,v,w;s)=uvw\,s-(uv+uw+vw)m_Q^2 ,
\label{eq:d0cutderive}
\end{equation}
where $K_{4}$ is the four-line kinematic polynomial whose sign determines the algebraic support of the truncated perturbative kernel. The Dirac traces generate a polynomial numerator in the loop momenta and masses. After the three shifted Gaussian momentum integrations, polynomial terms analytic in $p^2$ contribute only to subtraction terms, whereas logarithmic terms produce the physical discontinuity. Taking $\rho_0^{(2)}(s)=\pi^{-1}\operatorname{Im}\Pi_2^{(0)}(s+i0)$ across $K_{4}\ge0$ gives
\begin{equation}
\rho_0^{(2)}(s)=-\frac{64}{(4\pi)^6}
\int_0^1\!du\int_0^{1-u}\!dv\int_0^{1-u-v}\!dw\,
\frac{K_{4}^2}{u^3v^3w^3}\,
N_0^{(2)}(u,v,w;s,m_Q,m_s)\,\Theta(K_{4}),
\label{eq:d0derivedrho}
\end{equation}
where $N_0^{(2)}$ is the tensor-channel numerator polynomial given explicitly in Appendix \ref{app:full-symbolic}. The strange-quark mass is retained to first order in the corresponding Wilson coefficients.

The OPE representation satisfies a dispersion relation of the form
\begin{equation}
 \Pi_2^{\rm OPE}(p^2)=\int_{s_{\rm th}}^{\infty}\frac{\rho_2^{\rm OPE}(s)}{s-p^2}\,ds+\text{subtractions},
 \label{eq:dispersion}
\end{equation}
where the lower limit is taken as the quark-production threshold
\begin{equation}
 s_{\rm th}^{(Q)}=(3m_Q+m_s)^2,\qquad Q=c,b.
 \label{eq:sth}
\end{equation}
 Here, although the OPE Wilson coefficients are expanded through first order in $m_s$, the nonzero strange-quark mass is retained in $s_{\rm th}^{(Q)}$. Thus, the leading $m_s$ dependence is included both in the OPE and in the lower limit of the dispersion integral. Accordingly, applying the Borel transformation and the standard quark-hadron duality approximation for the continuum above $s_0$ yields the ground-state sum rule
\begin{equation}
 \lambda_{2^+}^{2}e^{-m_{2^+}^{2}/M_B^2}
 =\int_{s_{\rm th}}^{s_0}ds\,\rho_2^{\rm OPE}(s)e^{-s/M_B^2}.
 \label{eq:mastersr}
\end{equation}
where $M_B^2$ is the Borel parameter and $s_0$ is the continuum threshold separating the lowest-state contribution from the duality approximation to higher states. The Borel moments are defined by
\begin{equation}
  \mathcal L_n(M_B^2,s_0)=\int_{s_{\rm th}}^{s_0}\!ds\,s^{\,n}
  \rho_2^{\rm OPE}(s)e^{-s/M_B^2},
  \label{eq:moments}
\end{equation}
where $n$ denotes the moment order. Only $n=0$ and $n=1$ are required for the mass and pole residue. Consequently,
\begin{equation}
  m_{2^{+}}^2=\frac{\mathcal L_1}{\mathcal L_0},\qquad
  \lambda_{2^{+}}^2=\mathcal L_0\,e^{m_{2^+}^2/M_B^2}.
  \label{eq:massres}
\end{equation}
where $m_{2^+}$ is the extracted ground-state mass and $\lambda_{2^+}$ is its pole residue. The working region is constrained by the pole contribution and OPE convergence as
\begin{equation}
  PC=\frac{\mathcal L_0(M_B^2,s_0)}{\mathcal L_0(M_B^2,\infty)}\ge0.40,\qquad
  R_D=\frac{\mathcal L_0^{(D)}}{\mathcal L_0^{(0)}},
  \label{eq:criteria}
\end{equation}
where $PC$ is the pole contribution and $R_D$ measures the dimension-$D$ OPE contribution relative to the perturbative moment. The adopted convergence conditions are $|R_3|<1$ and $|R_D|<0.30$ for $D=4,5,6,7$, together with a stable Borel plateau.

Appendix \ref{app:full-symbolic} retains the complete perturbative numerator, the Wick contractions, propagator-insertion origin of the higher-dimensional sectors, all symbolic densities through $D\leq7$, and the distributional terms needed for numerical implementation.

\section{\texorpdfstring{Results and Discussion}{Results and Discussion}}
\label{sec:results}
\label{sec:numerics}

The charm and bottom channels are analyzed with the same acceptance criteria and the same sequence of numerical diagnostics. The condensate inputs are common to both channels, while the heavy-quark mass and the running coupling are evaluated at the corresponding heavy-quark scale.

\begin{table}[tb]
\centering
\caption{QCD input parameters used in the parallel $cc\bar c\bar s$ and $bb\bar b\bar s$ analyses. The central values and adopted variations are shown explicitly.}
\label{tab:inputs}
\begin{tabular}{lcc}
\toprule
Parameter & Central value & Adopted variation \\
\midrule
$m_c(m_c)$ & $1.273\GeV$ & $\pm0.005\GeV$ \\
$m_b(m_b)$ & $4.183\GeV$ & $\pm0.007\GeV$ \\
$m_s(2\GeV)$ & $0.0935\GeV$ & $\pm0.0008\GeV$ \\
$\langle\bar q q\rangle^{1/3}(1\GeV)$ & $-0.240\GeV$ & $\pm0.010\GeV$ \\
$\langle\bar s s\rangle/\langle\bar q q\rangle$ & $0.8$ & $\pm0.1$ \\
$m_0^2(1\GeV)$ & $0.8\GeVsq$ & $\pm0.1\GeVsq$ \\
$\langle g_s^2G^2\rangle$ & $0.47\ \mathrm{GeV}^4$ & $\pm0.14\ \mathrm{GeV}^4$ \\
$\alpha_s(m_c)$ & $0.39$ & $\pm0.05$ \\
$\alpha_s(m_b)$ & $0.224$ & $\pm0.008$ \\
$\kappa_s$ & $1.0$ & $1.0$-$2.0$ \\
\bottomrule
\end{tabular}
\end{table}

Table \ref{tab:inputs} collects the numerical inputs, which are used in the numerical computation. The heavy-quark and strange-quark masses are taken from \cite{ParticleDataGroup:2024cfk}. On the other hand,the quark, mixed, and gluon-condensate inputs follow the conventions commonly adopted
in QCD sum-rule analyses \cite{Reinders:1984sr,Colangelo:2000dp}.
The mixed condensates are defined by
\begin{equation}
 \langle\bar q g_s\sigma\!\cdot\!G q\rangle_{\mu_0}
 =m_0^2(\mu_0)\langle\bar q q\rangle_{\mu_0},\qquad
 \langle\bar s g_s\sigma\!\cdot\!G s\rangle_{\mu_0}
 =m_0^2(\mu_0)\langle\bar s s\rangle_{\mu_0},
 \qquad \mu_0=1\GeV,
 \label{eq:m0def}
\end{equation}
while the gluon-condensate convention is
\begin{equation}
 \langle g_s^2G^2\rangle
 =4\pi^2
 \left\langle\frac{\alpha_s}{\pi}G^2\right\rangle .
 \label{eq:GGconv}
\end{equation}
Furthermore, an explicit strong coupling appears in the factorized dimension six contribution,
which is proportional to
$\kappa_s g_s^2\langle\bar s s\rangle^2$. The parameter $\kappa_s$ accounts for deviations of the four-quark
condensate from the vacuum-saturation approximation
 \cite{Chetyrkin:2001je}. The factorized limit,
$\kappa_s=1$, is used for the central analysis, while
$1\leq\kappa_s\leq2$ is considered to assess the sensitivity to
possible violations of vacuum saturation.
Accordingly,
\begin{equation}
 g_s^2(\mu)=4\pi\alpha_s(\mu),
 \label{eq:gs_alpha}
\end{equation}
is evaluated at the corresponding heavy-quark scale,
$\mu=m_Q$ with $Q=c,b$. For the charm channel,
$\alpha_s(m_c)=0.39$ is adopted, whereas for the bottom channel
$\alpha_s(m_b)=0.224$ is used. These values are consistent with
determinations at the charm scale and with the standard
renormalization-group evolution of the strong coupling to the bottom
scale \cite{Maezawa:2016vgv,Beneke:2021lkq,ParticleDataGroup:2024cfk}.

\begin{figure}[!htbp]
  \centering
  \begin{minipage}[t]{0.485\textwidth}
    \centering
    \includecharmplot[width=\linewidth]{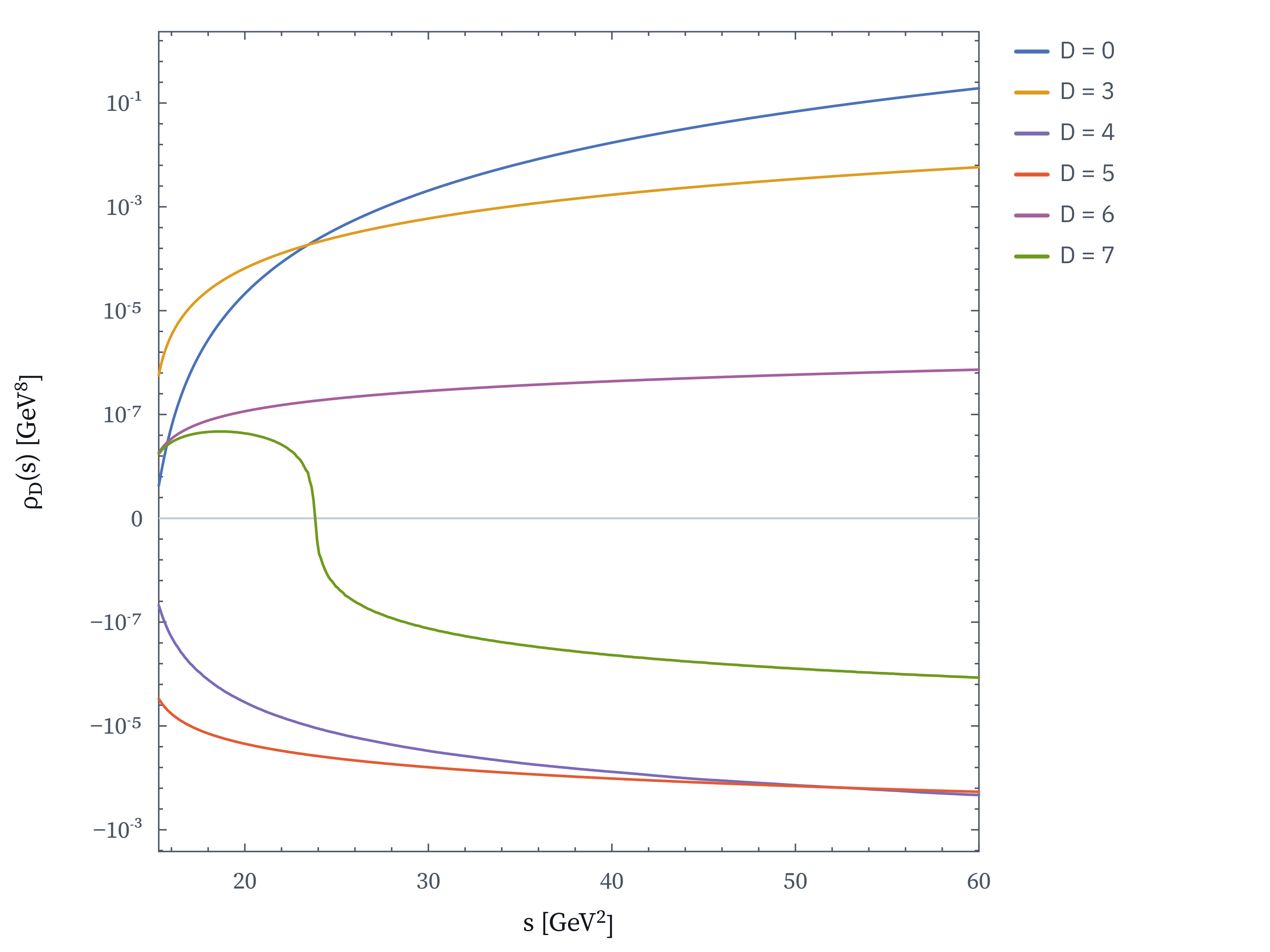}\\[-1pt]
    {\small (a) $cc\bar c\bar s$}
  \end{minipage}\hfill
  \begin{minipage}[t]{0.485\textwidth}
    \centering
    \includebottomplot[width=\linewidth]{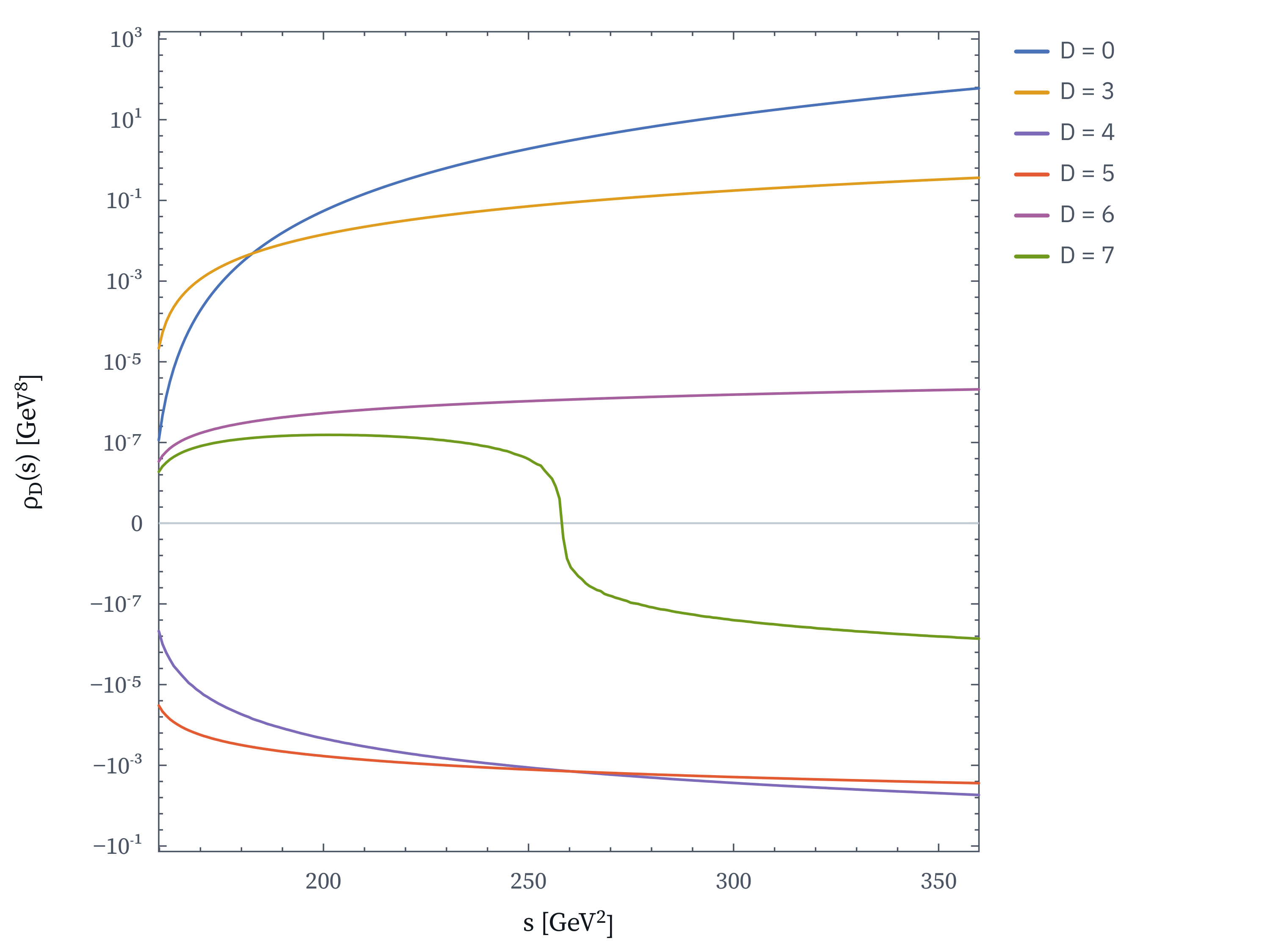}\\[-1pt]
    {\small (b) $bb\bar b\bar s$}
  \end{minipage}
  \caption{Continuous part of $\rho_D^{(2)}(s)$ for the retained OPE sectors. Panel (a) shows the $cc\bar c\bar s$ channel and panel (b) the $bb\bar b\bar s$ channel. Curves are labeled by the OPE dimension $D$.}
  \label{fig:densities}
\end{figure}

Figure \ref{fig:densities} is used as a first diagnostic of the relative hierarchy of the retained OPE sectors rather than as the convergence test itself. In both channels, the perturbative $D=0$ contribution grows most rapidly with $s$ and controls the spectral density at sufficiently large energy. The $D=3$ strange-quark condensate gives the leading positive nonperturbative correction, whereas the $D=4$ and $D=5$ sectors are negative over the displayed physical region. The $D=6$ and $D=7$ terms remain much smaller in magnitude than the leading sectors. The same qualitative ordering is therefore preserved under the heavy-flavour replacement $c\to b$, while the onset of the bottom spectral density is displaced to substantially larger $s$ by the heavier quark mass. Since only the continuous $\Theta$ pieces are displayed, the figure is not by itself used to judge OPE convergence. That requirement is imposed after Borel transformation through the moment ratios $R_D$.

\subsection{Borel windows, continuum thresholds, and OPE convergence}
\label{sec:window}

To determine the working windows, the same requirements are imposed in both channels as
\begin{equation}
 PC\ge0.40,\qquad |R_3|<1,\qquad |R_D|<0.30\quad(D=4,5,6,7).
 \label{eq:parallel-criteria}
\end{equation}
\begin{figure}[!htbp]
  \centering
  \begin{minipage}[t]{0.485\textwidth}
    \centering
    \includecharmplot[width=\linewidth]{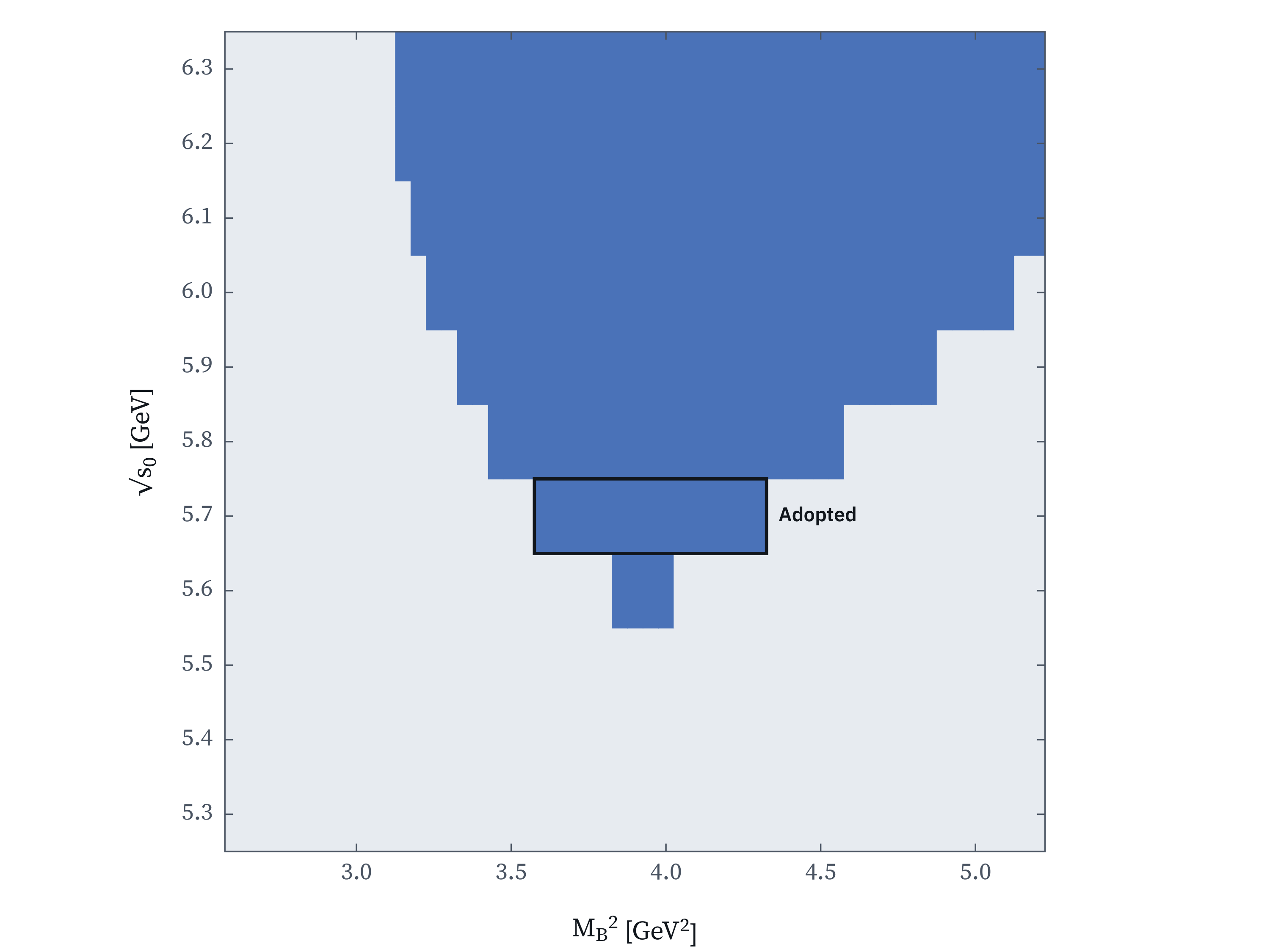}\\[-1pt]
    {\small (a) $cc\bar c\bar s$}
  \end{minipage}\hfill
  \begin{minipage}[t]{0.485\textwidth}
    \centering
    \includebottomplot[width=\linewidth]{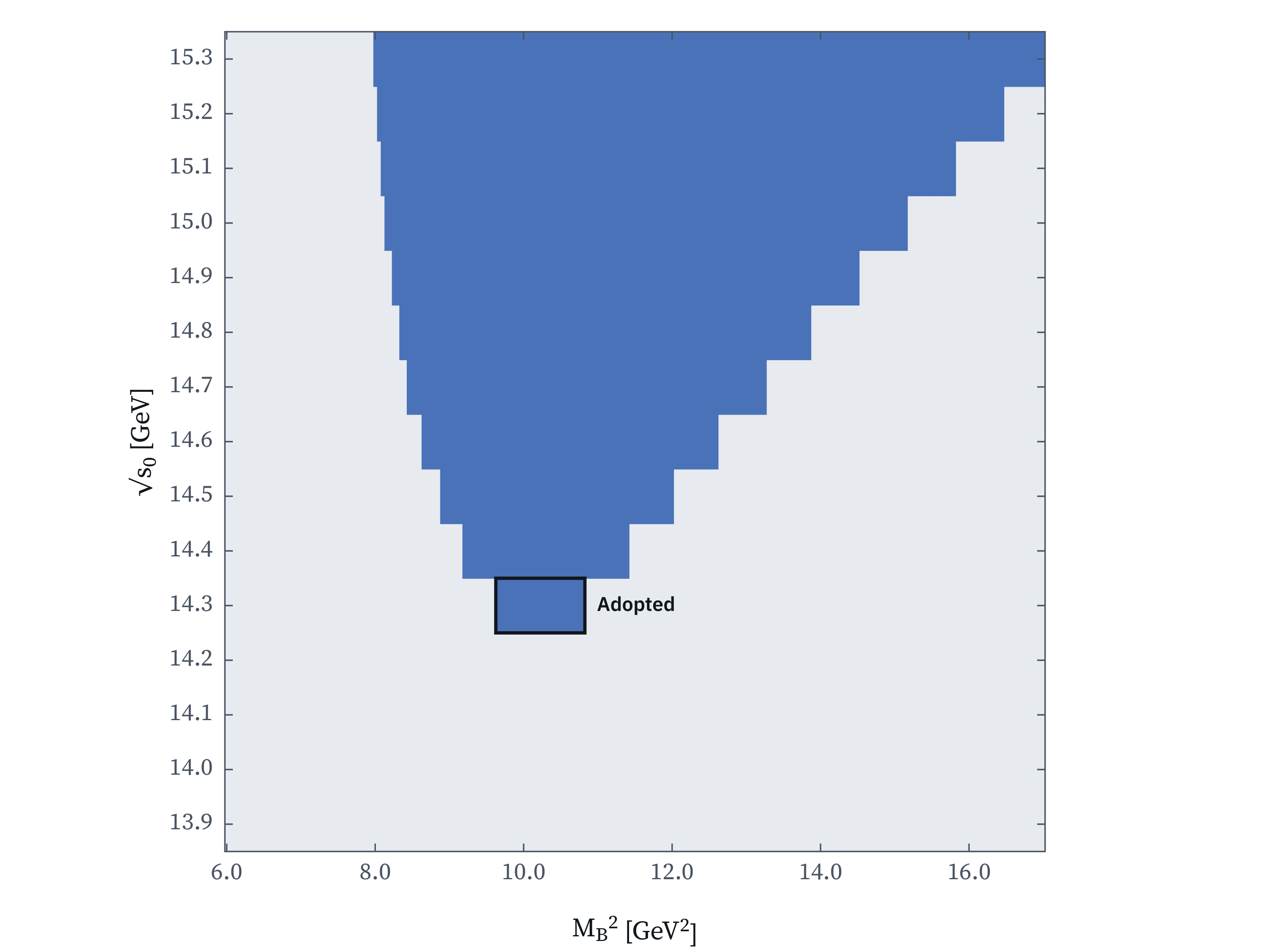}\\[-1pt]
    {\small (b) $bb\bar b\bar s$}
  \end{minipage}
  \caption{Accepted points in the $(M_B^2,\sqrt{s_0})$ plane. The outlined regions identify the working choices $\sqrt{s_0}=5.7\GeV$, $3.6\le M_B^2\le4.3\GeVsq$ in panel (a) and $\sqrt{s_0}=14.3\GeV$, $9.8\le M_B^2\le10.8\GeVsq$ in panel (b).}
  \label{fig:region}
\end{figure}

The accepted regions in Fig. \ref{fig:region} are obtained from the simultaneous pole-dominance and OPE-convergence requirements, so the lower and upper Borel boundaries have different origins. At small $M_B^2$, higher-dimensional terms become comparatively more important and set the convergence limit, whereas at large $M_B^2$ the pole contribution decreases and eventually reaches the $40\%$ floor. In the charm channel, the adopted $\sqrt{s_0}=5.7\GeV$ choice gives $0.7\GeVsq$ working interval for the Borel parameter. In the bottom channel, $\sqrt{s_0}=14.3\GeV$ already provides a $1.0\GeVsq$ interval, $9.8$-$10.8\GeVsq$. Accordingly, the selected regions are
\begin{align}
 cc\bar c\bar s:&\qquad \sqrt{s_0}=5.7\GeV,\qquad 3.6\le M_B^2\le4.3\GeVsq,\label{eq:windowc}\\
 bb\bar b\bar s:&\qquad \sqrt{s_0}=14.3\GeV,\qquad 9.8\le M_B^2\le10.8\GeVsq.\label{eq:windowb}
\end{align}
On the other hand, the lower dispersion limits are $s_{\rm th}^{(c)}=(3m_c+m_s)^2\simeq15.31\GeVsq$ and $s_{\rm th}^{(b)}=(3m_b+m_s)^2\simeq159.83\GeVsq$.

\begin{figure}[!htbp]
  \centering
  \begin{minipage}[t]{0.485\textwidth}
    \centering
    \includecharmplot[width=\linewidth]{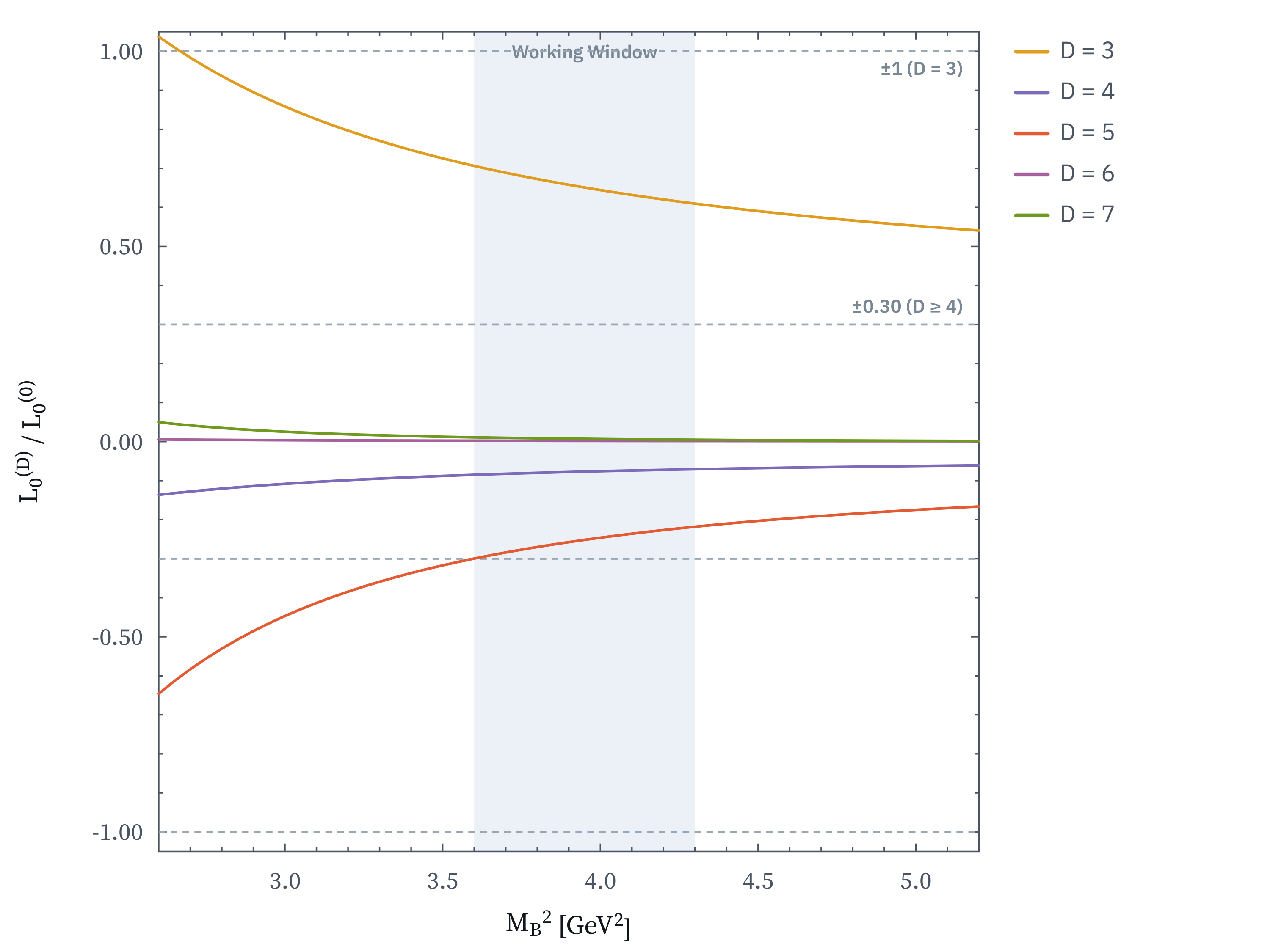}\\[-1pt]
    {\small (a) $cc\bar c\bar s$}
  \end{minipage}\hfill
  \begin{minipage}[t]{0.485\textwidth}
    \centering
    \includebottomplot[width=\linewidth]{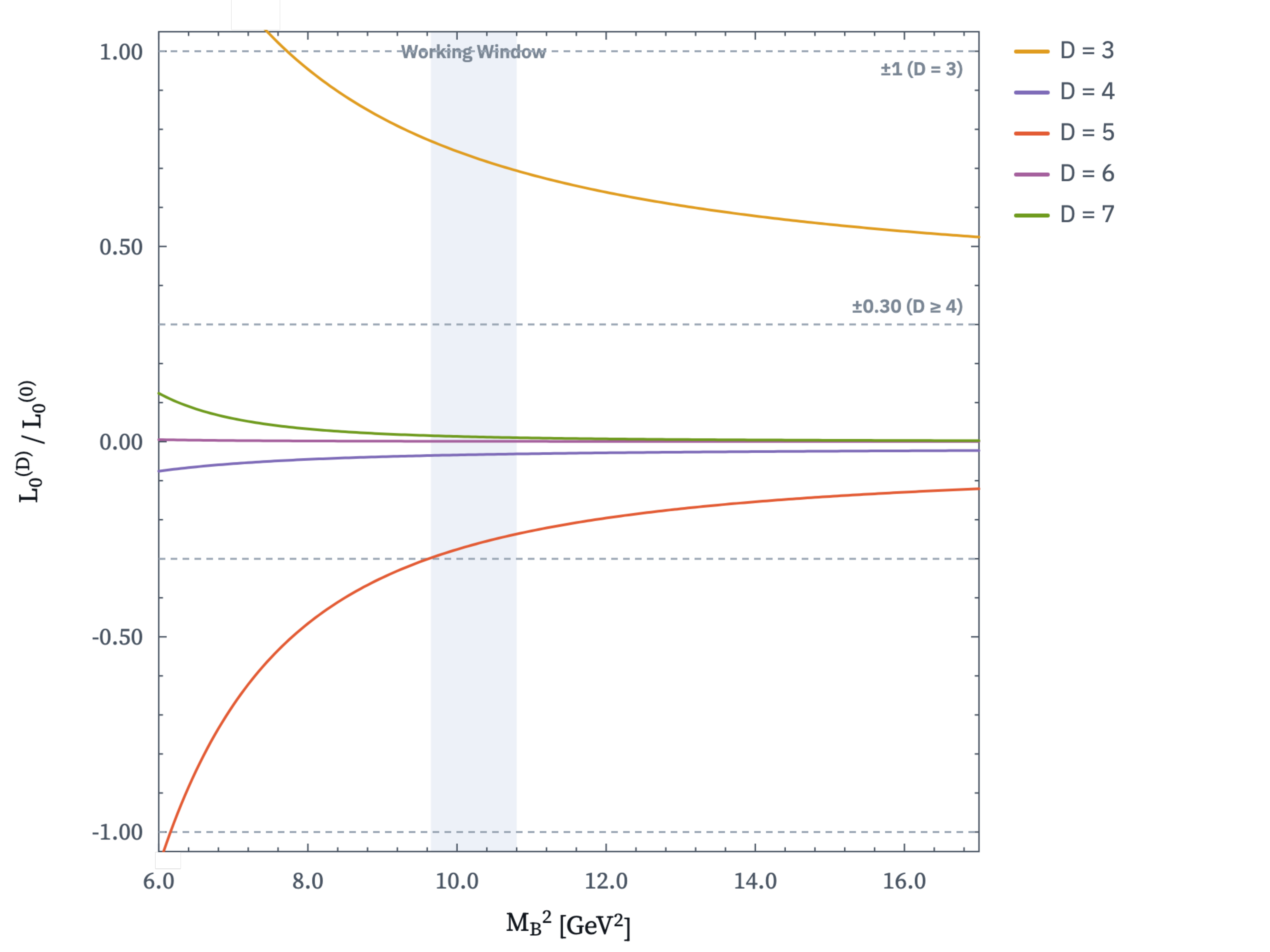}\\[-1pt]
    {\small (b) $bb\bar b\bar s$}
  \end{minipage}
  \caption{Signed OPE ratios $R_D=\mathcal L_0^{(D)}/\mathcal L_0^{(0)}$ as functions of $M_B^2$. The curves correspond to $D=3,4,5,6,7$, shaded bands mark the working Borel intervals and horizontal reference lines show the adopted convergence bounds.}
  \label{fig:ope}
\end{figure}

\begin{figure}[!htbp]
  \centering
  \begin{minipage}[t]{0.485\textwidth}
    \centering
    \includecharmplot[width=\linewidth]{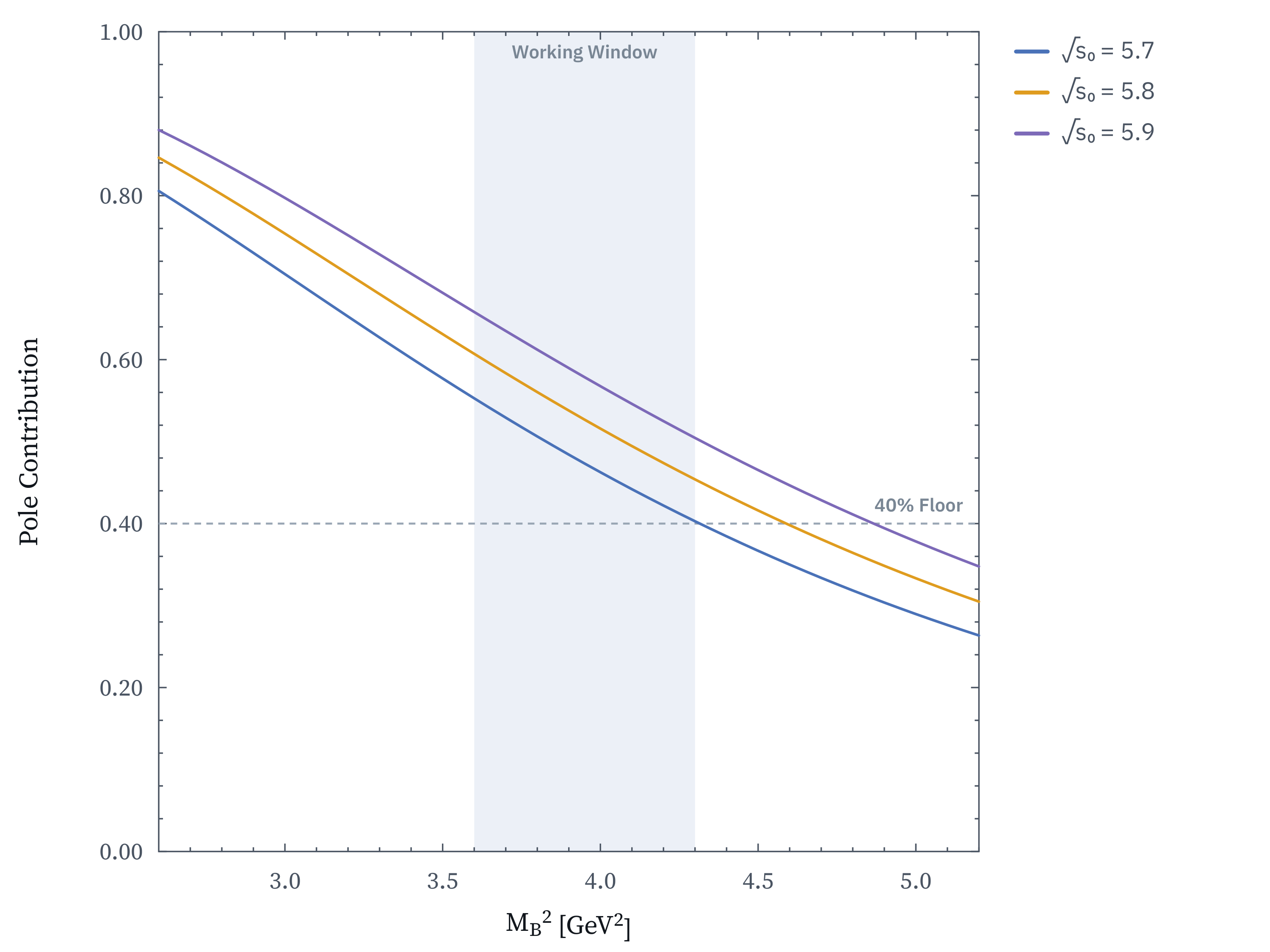}\\[-1pt]
    {\small (a) $cc\bar c\bar s$}
  \end{minipage}\hfill
  \begin{minipage}[t]{0.485\textwidth}
    \centering
    \includebottomplot[width=\linewidth]{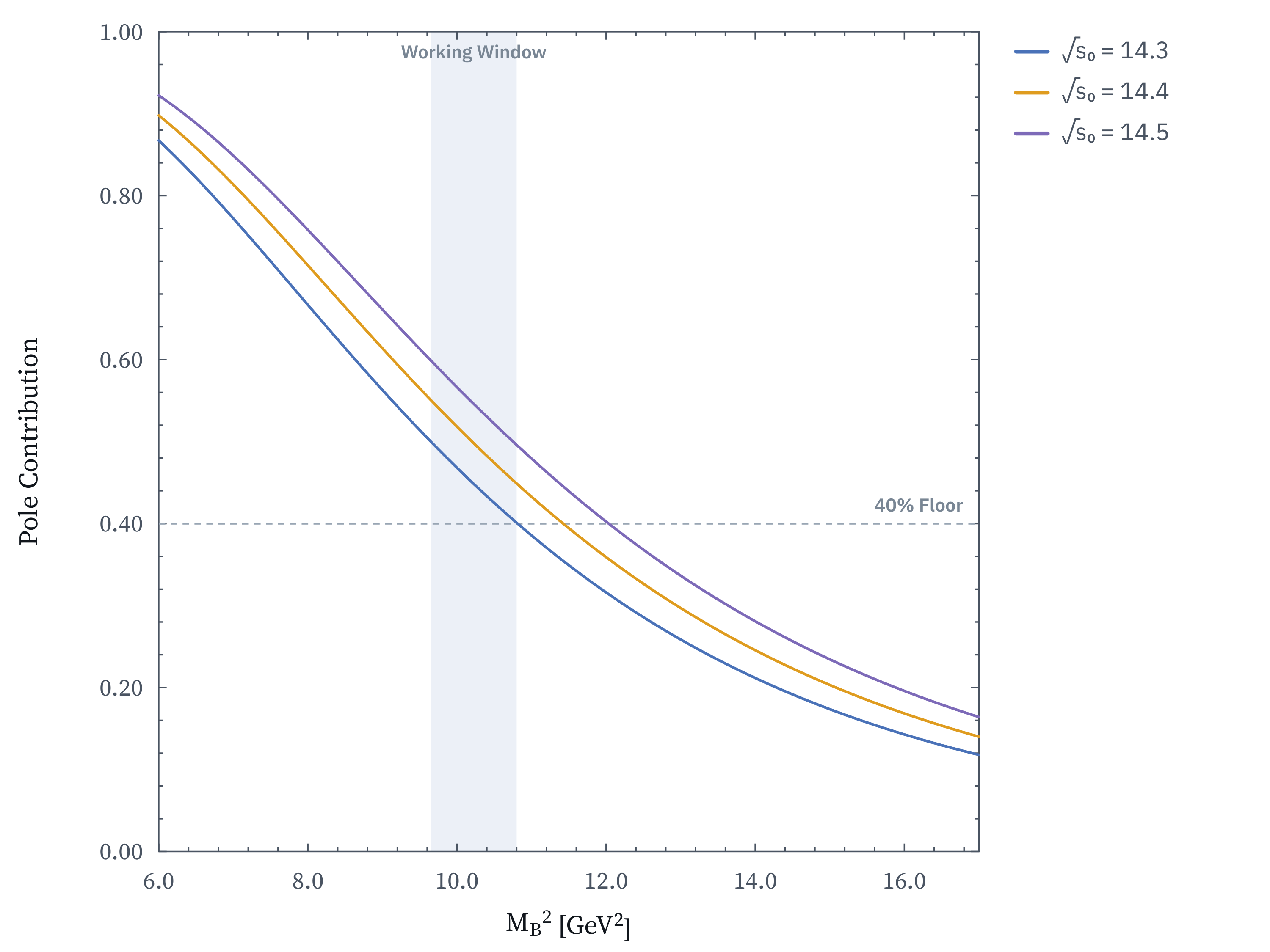}\\[-1pt]
    {\small (b) $bb\bar b\bar s$}
  \end{minipage}
  \caption{Pole contribution $PC$ as a function of $M_B^2$ for representative continuum thresholds. The horizontal line marks $PC=0.40$, and the shaded bands mark the working Borel intervals.}
  \label{fig:pc}
\end{figure}

At the representative Borel points $M_B^2=4.0\GeVsq$ and $10.3\GeVsq$, the OPE ratios are
\begin{align}
cc\bar c\bar s:&\quad R_3=0.644,\quad R_4=-0.0761,\quad R_5=-0.246,\quad R_6=0.00175,\quad R_7=0.00666,\label{eq:ratiosc}\\
bb\bar b\bar s:&\quad R_3=0.723,\quad R_4=-0.0330,\quad R_5=-0.268,\quad R_6=0.000814,\quad R_7=0.0124.\label{eq:ratiosb}
\end{align}
Thus, the same convergence bounds are satisfied at the central points of both windows. Figures \ref{fig:ope} and \ref{fig:pc} display the corresponding Borel dependence of the OPE ratios and pole contribution.

Figure \ref{fig:ope} makes the convergence pattern more explicit. The $D=3$ condensate is the largest nonperturbative correction in both channels, with central ratios $0.644$ for charm and $0.723$ for bottom. The $D=5$ mixed-condensate term provides the largest negative correction as $-0.268$, and partially compensates the $D=3$ contribution. The $D=4$ term is also negative but is smaller in magnitude, particularly in the bottom channel, while the $D=6$ and $D=7$ ratios are at the percent level or below. Across both shaded working windows, all curves remain within the bounds of Eq. \eqref{eq:parallel-criteria}. The decrease of the absolute higher-dimensional ratios with increasing $M_B^2$ also explains why OPE convergence controls the lower edge of the allowed region. The small retained $D=6$ and $D=7$ terms support numerical convergence within the declared operator basis, although they do not quantify the omitted independent $\langle g_s^3G^3\rangle$ contribution.

The pole-contribution curves in Fig. \ref{fig:pc} decrease monotonically with $M_B^2$ for fixed $s_0$, while a larger continuum threshold increases the pole fraction at the same Borel point. This behaviour determines the upper edge of each Borel window. At the representative points, the pole contributions are $46.3\%$ for $cc\bar c\bar s$ and $44.1\%$ for $bb\bar b\bar s$, so both are safely above the adopted $40\%$ requirement. In the bottom window the pole contribution decreases from approximately $48.5\%$ to $40.0\%$ between $M_B^2=9.8$ and $10.8\GeVsq$; the upper endpoint is therefore fixed essentially by the pole-dominance criterion. The charm panel exhibits the same competition between a convergent OPE at larger $M_B^2$ and a decreasing pole fraction, which is the reason the accepted window remains finite even after the OPE ratios have become smaller.

\subsection{Mass, pole residue, and uncertainty analysis}
\label{sec:mass}

At the representative points in the two windows, the extracted quantities are
\begin{align}
cc\bar c\bar s:\quad &M_B^2=4.0\GeVsq, &m_{2^+}&=5.239\GeV, &\lambda_{2^+}&=0.113\ \mathrm{GeV}^5, &PC&=46.3\%,\label{eq:centralc}\\
bb\bar b\bar s:\quad &M_B^2=10.3\GeVsq, &m_{2^+}&=13.832\GeV, &\lambda_{2^+}&=0.851\ \mathrm{GeV}^5, &PC&=44.1\%.\label{eq:centralb}
\end{align}
Across the adopted windows, the charm mass varies from $5.216$ to $5.254\GeV$, while the bottom mass varies from $13.822$ to $13.842\GeV$. The latter has a mean of $13.832\GeV$ and a standard deviation of about $0.006\GeV$ over the sampled window. For the bottom channel, the pole contribution decreases from approximately $48.5\%$ at the lower edge to $40.0\%$ at the upper edge.

\begin{figure}[!htbp]
  \centering
  \begin{minipage}[t]{0.485\textwidth}
    \centering
    \includecharmplot[width=\linewidth]{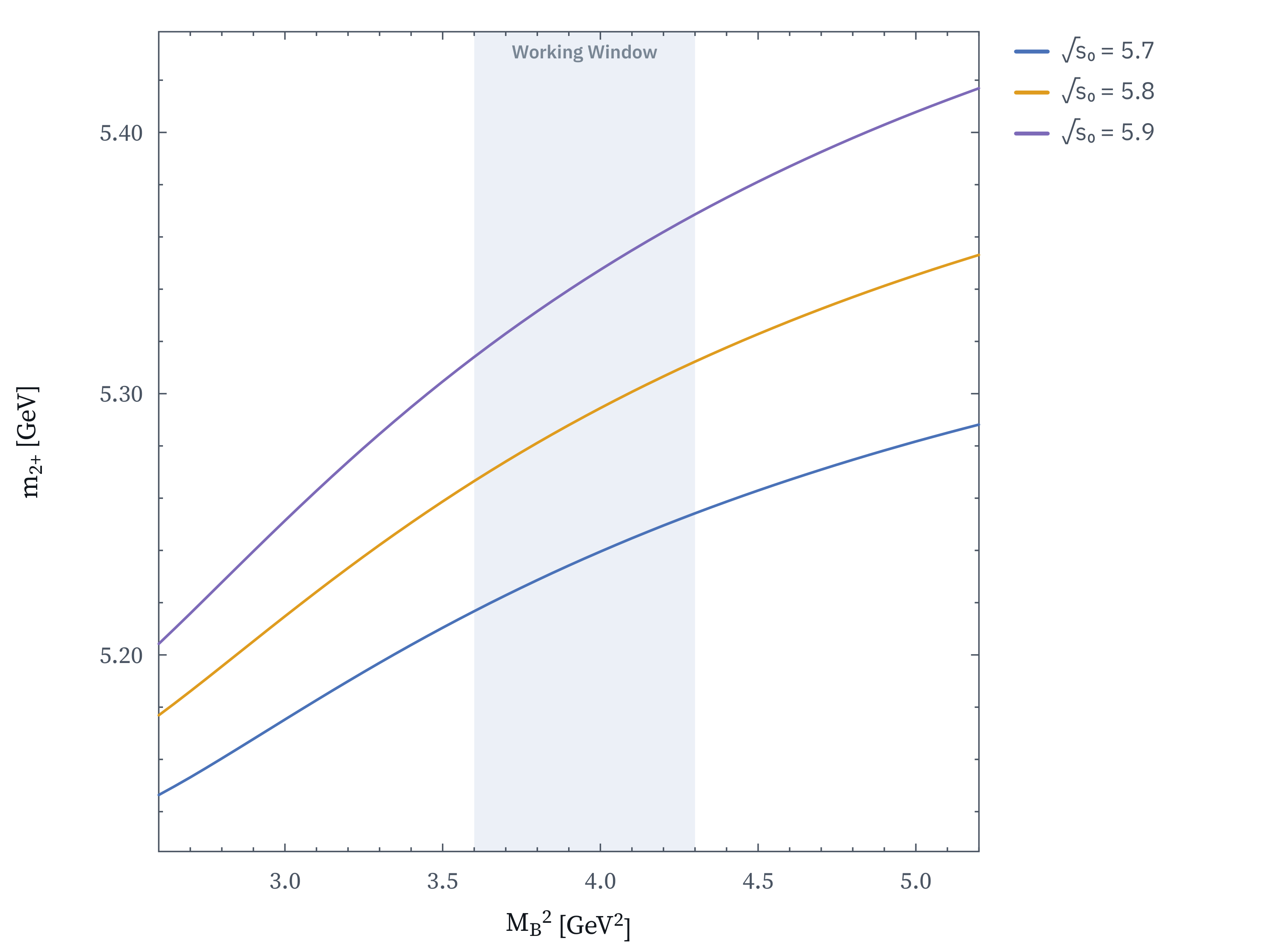}\\[-1pt]
    {\small (a) $cc\bar c\bar s$}
  \end{minipage}\hfill
  \begin{minipage}[t]{0.485\textwidth}
    \centering
    \includebottomplot[width=\linewidth]{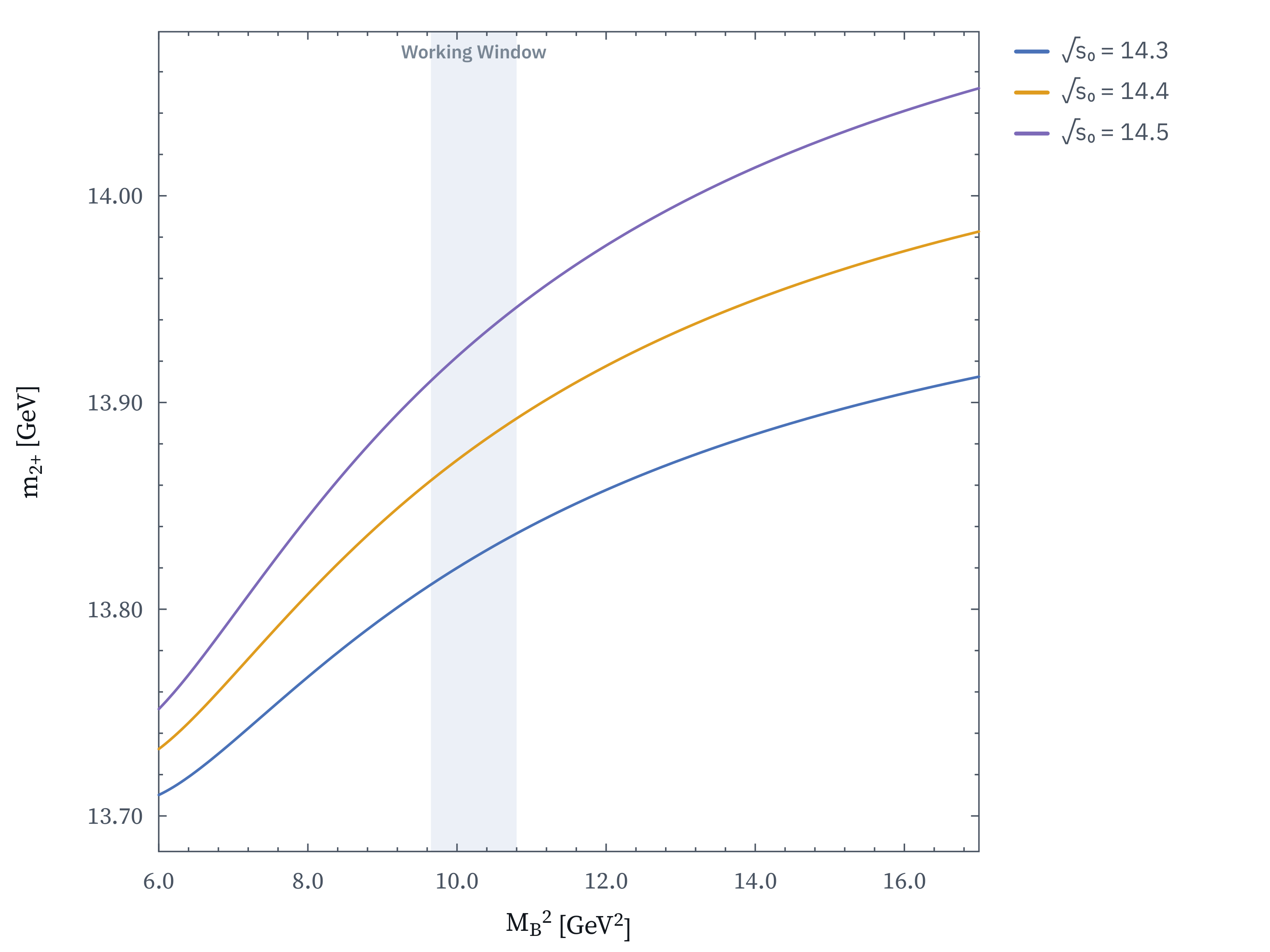}\\[-1pt]
    {\small (b) $bb\bar b\bar s$}
  \end{minipage}
  \caption{Extracted mass $m_{2^+}$ as a function of $M_B^2$ for representative continuum thresholds $\sqrt{s_0}$. The shaded bands mark the working Borel intervals in the charm and bottom channels.}
  \label{fig:mass}
\end{figure}

Figure \ref{fig:mass} shows that the extracted mass changes smoothly with $M_B^2$ and increases with the continuum threshold in both channels. Inside the adopted windows, however, the residual Borel dependence is modest: the charm result spans $5.216$-$5.254\GeV$, a range of about $38\MeV$, whereas the bottom result spans $13.822$-$13.842\GeV$, about $20\MeV$. Relative to the central masses, these ranges correspond to approximately $0.7\%$ and $0.14\%$, respectively. The bottom sum rule therefore exhibits the flatter mass behaviour within its accepted interval. In both panels, the separation among the different $\sqrt{s_0}$ curves is more important than the drift inside a single working window, anticipating the dominant role of the continuum-threshold variation in the final uncertainty budget.

Table \ref{tab:parallel-summary} summarizes the quantities used for the direct charm-bottom comparison. Moreover, Table \ref{tab:budget} lists the representative half-range mass shifts used in the quadrature budget for each states.
\begin{table}[tb]
\centering
\caption{Parallel numerical summary for the two spin-2 channels.}
\label{tab:parallel-summary}
\begin{tabular}{lcc}
\toprule
Quantity & $cc\bar c\bar s$ & $bb\bar b\bar s$ \\
\midrule
$s_{\rm th}$ ($\mathrm{GeV}^2$) & $15.31$ & $159.83$ \\
$\sqrt{s_0}$ (GeV) & $5.7$ & $14.3$ \\
$M_B^2$ window ($\mathrm{GeV}^2$) & $3.6$-$4.3$ & $9.8$-$10.8$ \\
Representative $M_B^2$ ($\mathrm{GeV}^2$) & $4.0$ & $10.3$ \\
$m_{2^+}$ at representative point (GeV) & $5.239$ & $13.832$ \\
$\lambda_{2^+}$ ($\mathrm{GeV}^5$) & $0.113$ & $0.851$ \\
$PC$ at representative point & $46.3\%$ & $44.1\%$ \\
$R_3$ & $0.644$ & $0.723$ \\
$R_4$ & $-0.0761$ & $-0.0330$ \\
$R_5$ & $-0.246$ & $-0.268$ \\
$R_6$ & $0.00175$ & $0.000814$ \\
$R_7$ & $0.00666$ & $0.0124$ \\
Final mass (GeV) & $5.239\pm0.061$ & $13.832\pm0.059$ \\
\bottomrule
\end{tabular}
\end{table}

\begin{table}[tb]
\centering
\caption{Representative half-range mass shifts used in the uncertainty budgets. All entries are in MeV.}
\label{tab:budget}
\begin{tabular}{lcc}
\toprule
Source & $cc\bar c\bar s$ & $bb\bar b\bar s$ \\
\midrule
$\sqrt{s_0}$ & $55.2$ & $53.9$ \\
$M_B^2$ window & $18.7$ & $12.4$ \\
$m_0^2$ & $15.5$ & $16.3$ \\
$m_Q$ & $5.7$ & $9.2$ \\
$\langle\bar s s\rangle/\langle\bar q q\rangle$ & $4.1$ & $4.3$ \\
$\langle\bar q q\rangle^{1/3}$ & $4.0$ & $4.3$ \\
$\langle g_s^2G^2\rangle$ & $3.6$ & $0.9$ \\
$\kappa_s$ & $0.5$ & $0.3$ \\
$\alpha_s$ & $0.1$ & $0.1$ \\
$m_s$ & $0.1$ & $0.1$ \\
\midrule
QCD-input quadrature & $17.9$ & $19.7$ \\
Total quadrature & $61.0$ & $58.7$ \\
\bottomrule
\end{tabular}
\end{table}

\begin{figure}[!htbp]
  \centering
  \begin{minipage}[t]{0.485\textwidth}
    \centering
    \includecharmplot[width=\linewidth]{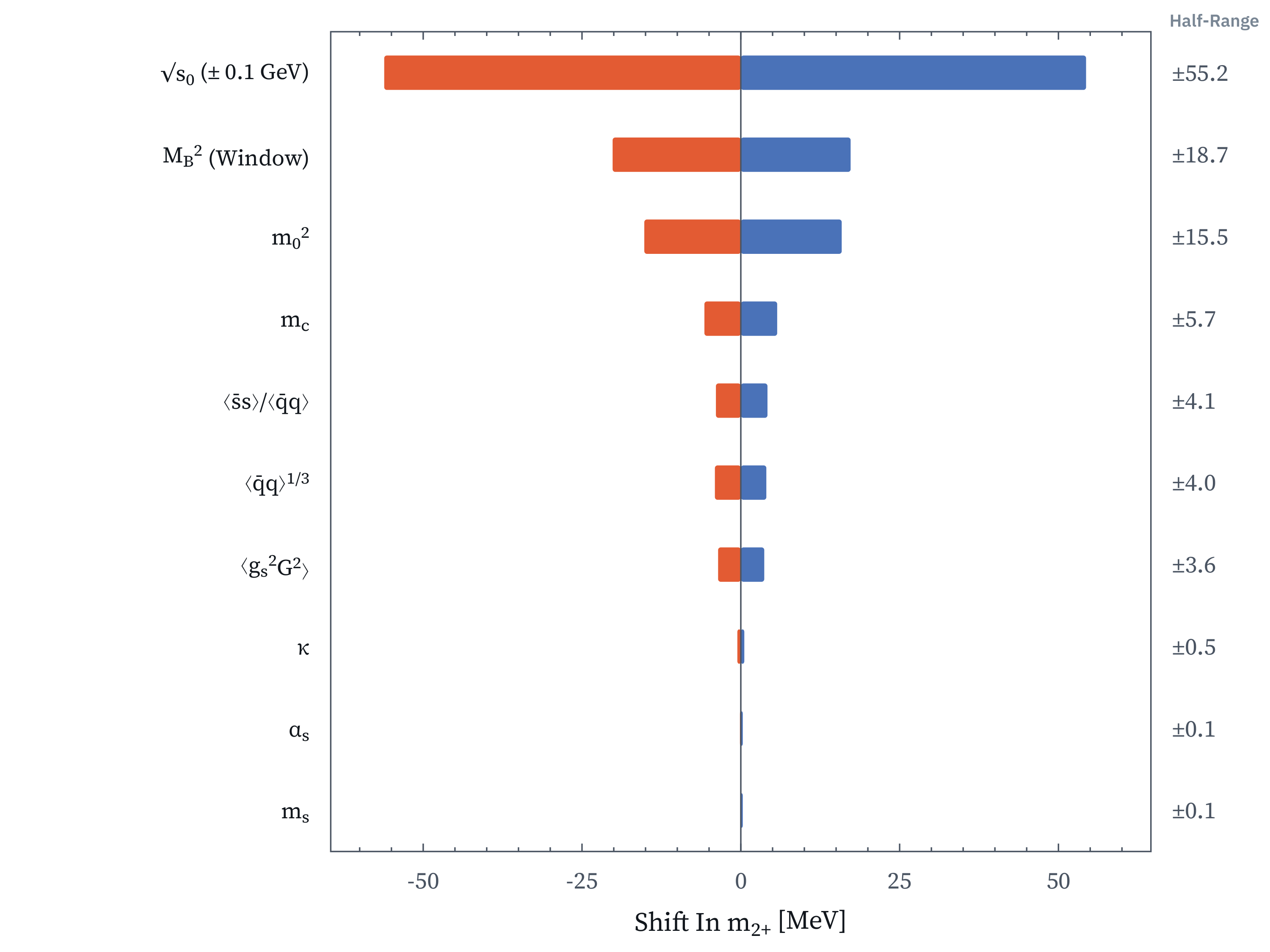}\\[-1pt]
    {\small (a) $cc\bar c\bar s$}
  \end{minipage}\hfill
  \begin{minipage}[t]{0.485\textwidth}
    \centering
    \includebottomplot[width=\linewidth]{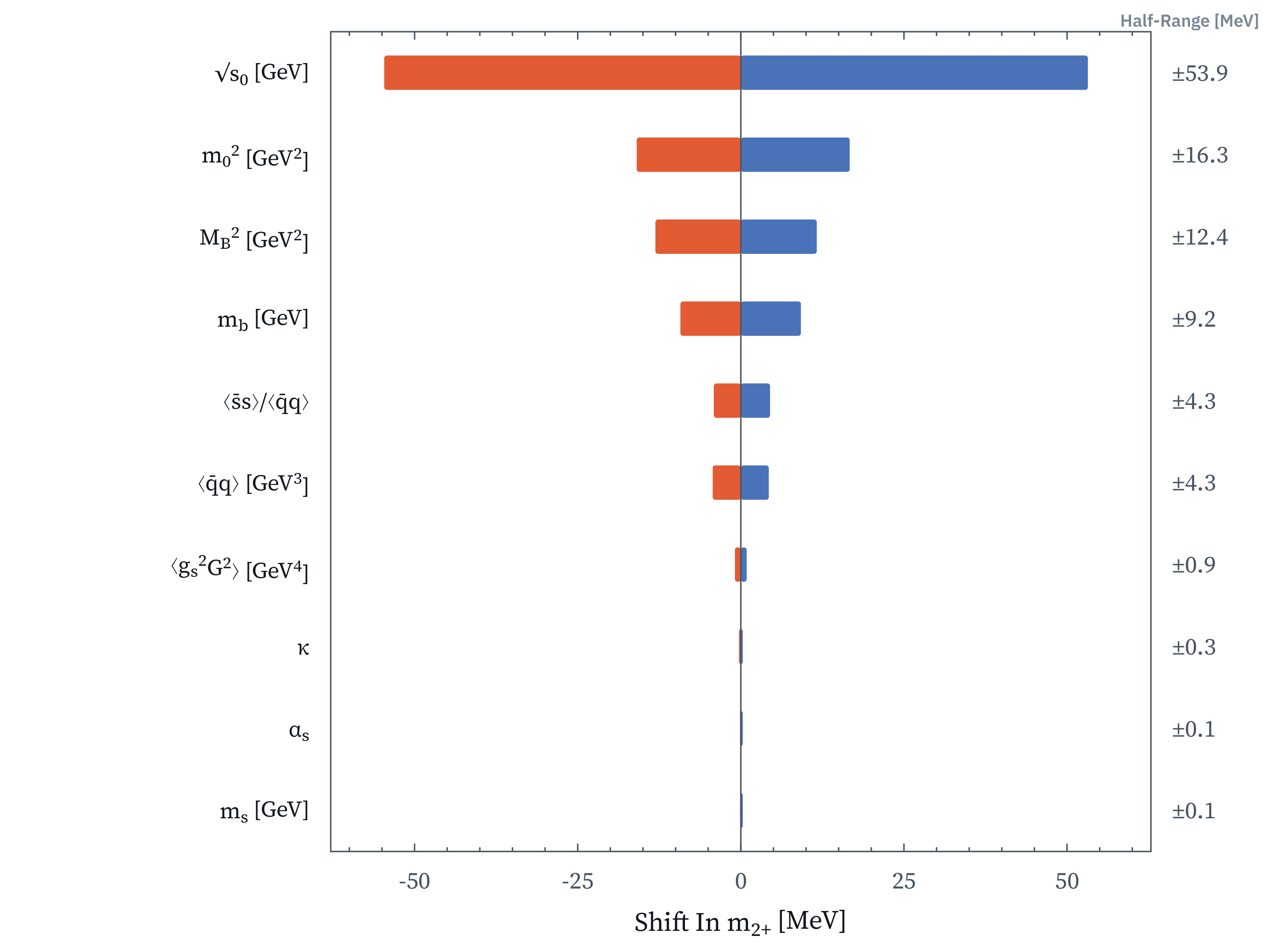}\\[-1pt]
    {\small (b) $bb\bar b\bar s$}
  \end{minipage}
  \caption{Mass-uncertainty contributions from the individual QCD inputs and auxiliary parameters. Horizontal bars show the half-range mass shifts in MeV used in the charm and bottom uncertainty budgets.}
  \label{fig:tornado}
\end{figure}

The comparison in Fig. \ref{fig:tornado} shows that the uncertainty hierarchy is similar, but not identical, in the two channels. The continuum threshold is the dominant source in both cases, producing half-range shifts of $55.2\MeV$ for charm and $53.9\MeV$ for bottom. The Borel-window contribution is larger for charm, $18.7\MeV$ versus $12.4\MeV$, whereas the mixed-condensate parameter gives comparable shifts of $15.5\MeV$ and $16.3\MeV$. The heavy-quark mass uncertainty is more visible in the bottom channel, increasing from $5.7\MeV$ for $m_c$ to $9.2\MeV$ for $m_b$, while the gluon-condensate sensitivity decreases from $3.6\MeV$ to $0.9\MeV$. Consequently, the QCD-input quadrature is slightly larger for bottom, $19.7\MeV$ compared with $17.9\MeV$, but the smaller Borel-window contribution makes its full quadrature uncertainty slightly smaller. This distinction is obscured if both errors are rounded immediately to $0.06\GeV$.

The quadrature totals are therefore $61.0\MeV$ for the charm channel and $58.7\MeV$ for the bottom channel. To retain this small but genuine difference between the propagated uncertainties, the final results are quoted with the corresponding MeV-level precision as
\begin{equation}
 m_{cc\bar c\bar s,\,2^+}=5.239\pm0.061\GeV,\qquad
 m_{bb\bar b\bar s,\,2^+}=13.832\pm0.059\GeV.
 \label{eq:finalmasses}
\end{equation}
The continuum-threshold variation is the largest individual contribution in both channels. The mixed-condensate parameter and Borel-window variation give the next largest contributions, while the $m_s$, $\alpha_s$, and $\kappa_s$ variations are numerically small on the scale of the final uncertainties.

\section{Conclusion}

For the charm channel, the central value in Eq. \eqref{eq:finalmasses} is approximately $0.25\GeV$ below the $5.489\GeV$ NRQM result for the same $J^P=2^+$ flavour configuration \cite{An:2026TriplyHeavy}. The $J/\psi D_s^*$ threshold obtained from $m_{J/\psi}=3.0969\GeV$ and $m_{D_s^*}=2.1122\GeV$ is $5.2091\GeV$ \cite{ParticleDataGroup:2024cfk}. The central QCD Sum Rules value is therefore about $31\MeV$ above this threshold, while the quoted uncertainty overlaps it. Consequently, the mass sum rule alone does not determine on which side of the threshold the physical state lies.

For the bottom channel, Ref. \cite{An:2026TriplyHeavy} predicts a $J^P=2^+$ $bb\bar b\bar s$ mass of $15.238\GeV$, which is approximately $1.41\GeV$ above the QCD Sum Rules central value in Eq. \eqref{eq:finalmasses}. The relevant S-wave rearrangement threshold for a spin-2 state is $\Upsilon B_s^*$, since the two vector mesons can couple to total $J=2$ in an S wave. Using $m_{\Upsilon}=9.4603\GeV$ and $m_{B_s^*}=5.4154\GeV$, the threshold is $14.8757\GeV$ \cite{ParticleDataGroup:2024cfk}. The central QCD Sum Rules value therefore lies about $1.04\GeV$ below this threshold. It also lies below the lower $\eta_b B_s$ two-body threshold, $14.7648\GeV$, by about $0.93\GeV$. Hence, unlike the charm case, the quoted parametric uncertainty does not overlap the open bottomonium-$B_s^{(*)}$ rearrangement thresholds. Within the present sum-rule setup, the $bb\bar b\bar s$ state is therefore predicted below these two-body thresholds, subject to the theoretical systematics stated below.

Taken together, the parallel analysis gives $m_{cc\bar c\bar s,\,2^+}=5.239\pm0.061\GeV$ and $m_{bb\bar b\bar s,\,2^+}=13.832\pm0.059\GeV$. At the representative Borel points, the pole contributions are about $46\%$ and $44\%$, respectively, while the retained $D=6$ and $D=7$ terms remain small compared with the leading OPE sectors. The continuum-threshold variation is the largest individual contribution to both uncertainty budgets. The total propagated uncertainties, $61.0\MeV$ for charm and $58.7\MeV$ for bottom, are therefore similar in magnitude but arise from slightly different numerical sensitivities, as shown in Fig. \ref{fig:tornado} and Table \ref{tab:budget}.

The quoted uncertainties include the adopted QCD-input ranges, the continuum-threshold variation, and the variation within the accepted Borel windows. They should not be interpreted as complete estimates of every theoretical systematic. Alternative tensor currents, higher-order Wilson coefficients, renormalization-scale dependence beyond the adopted prescription, and the independent triple-gluon condensate $\langle g_s^3G^3\rangle$ remain outside the stated numerical errors. The strange-quark mass is retained through first order in the Wilson coefficients and in $s_{\rm th}^{(Q)}=(3m_Q+m_s)^2$. Within these stated assumptions, the two channels provide a consistent heavy-flavour pair within the same spin-2 QCD Sum Rules construction.

\FloatBarrier

\section*{Acknowledgments}
The author acknowledges YEKUT, the High
Energy and Quantum Computations Laboratory at Yozgat Bozok University, where the computational part
of this study was carried out.

%\section*{Data availability}
%The data supporting the findings of this study are available from the author upon reasonable request.

%\section*{Code availability}
%The numerical codes used for the Borel-window, pole-contribution, mass, and pole-residue analyses are available from the author upon reasonable request. Parts of the current-specific symbolic calculation files may also be provided upon reasonable request.

\appendix
\section{Appendix}
\label{app:full-symbolic}

This appendix contains the detailed current-specific expressions that complement the main mass-extraction narrative. The current and the retained OPE decomposition are already defined in Eqs. \eqref{eq:current} and \eqref{eq:rhoope-decomp}, respectively, and are not repeated here. The symbolic results below are written in terms of the generic heavy-quark mass $m_Q$, with $Q=c$ or $b$; the bottom expressions therefore follow from the same formulas without duplicating the Appendix. The symbolic results below use a manuscript-specific simple Feynman-parameter notation. The change of variables is purely algebraic and does not alter any Wilson coefficient or integration domain. Accordingly, Ref. \cite{Xu:2025} is retained only as a scalar-channel comparison, not as the source of the notation used below.

\subsection{\texorpdfstring{Feynman parameters and integration domains}{Feynman parameters and integration domains}}
{
For the four-line representation, the Feynman parameters $u,v,w$ are used with
\begin{equation}
t_4=1-u-v-w,\qquad
K_{2}(s)=uv\,s-(u+v)m_Q^2,
\label{eq:app-t4-k2}
\end{equation}
where $t_4$ is the fourth Feynman parameter and $K_{2}$ is the reduced pair denominator. The four-line kinematic polynomial $K_{4}$ is the one already defined in Eq. \eqref{eq:d0cutderive}. For reduced two-fold representations,
\begin{equation}
t_3=1-u-v,\qquad
K_{3}(u,v;s)=uv\,t_3\,s-[uv+u t_3+v t_3]m_Q^2,
\label{eq:app-t3-k3}
\end{equation}
where $t_3$ is the remaining reduced Feynman parameter and $K_{3}$ is the corresponding three-line kinematic polynomial. The physical-cut boundaries are
\begin{align}
u_{\pm}&=\frac{s-3m_Q^2\pm\sqrt{(s-3m_Q^2)^2-4sm_Q^2}}{2s},\\
v_{\pm}(u)&=\frac{1-u\pm\sqrt{(1-u)^2-\dfrac{u(1-u)m_Q^2}{us-m_Q^2}}}{2},\\
w_0(u,v)&=\frac{uvm_Q^2}{K_{2}(s)}.
\label{eq:app-cut-bounds}
\end{align}
The following domain shorthand is used:
\begin{align}
\int_{R_4(s)}&\equiv\int_{u_-}^{u_+}\dd u\int_{v_-(u)}^{v_+(u)}\dd v\int_{w_0(u,v)}^{1-u-v}\dd w,\\
\int_{R_3(s)}&\equiv\int_{u_-}^{u_+}\dd u\int_{v_-(u)}^{v_+(u)}\dd v,\\
\int_{S_4}&\equiv\int_0^1\dd u\int_0^{1-u}\dd v\int_0^{1-u-v}\dd w,\\
\int_{S_3}&\equiv\int_0^1\dd u\int_0^{1-u}\dd v.
\label{eq:app-domains}
\end{align}
Here $R_4$ and $R_3$ denote the physical integration regions, while $S_4$ and $S_3$ denote the corresponding full parameter domains.
}

\subsection{\texorpdfstring{$D=0$}{D=0: explicit numerator and normalization checks}}

The Wick contraction and step-by-step perturbative reduction are given in Sections \ref{subsec:current-wick} and \ref{subsec:d0-method}. The fully expanded numerator polynomial and compact normalization checks are retained here.

% ===========================================================================

The kinematic quantities and integration domains are those of Eqs. \eqref{eq:d0cutderive}, \eqref{eq:app-t3-k3}, and \eqref{eq:app-domains}. The perturbative spectral density itself is already given in Eq. \eqref{eq:d0derivedrho}; only its explicit numerator is retained here. The polynomial $N_0^{(2)}$ has homogeneous mass dimension four and is symmetric under $u\leftrightarrow v$. The numerator polynomial is fully explicit:
\begin{align}
&N_0^{(2)}=u^{2} v^{3} m_Q^{4} + u^{2} w^{3} m_Q^{4} + u^{3} v^{2} m_Q^{4} + u^{3} w^{2} m_Q^{4} + v^{2} w^{3} m_Q^{4} + v^{3} w^{2} m_Q^{4} - u^{2} v^{2} m_Q^{4} - 3 u^{2} w^{2} m_Q^{4} \\ \nonumber
&\qquad - 3 v^{2} w^{2} m_Q^{4} - 2 u w^{3} m_Q^{4} - 2 v w^{3} m_Q^{4} + 2 u w^{2} m_Q^{4} + 2 v w^{2} m_Q^{4} - 15 u^{2} v^{2} w^{2} s^{2} - 8 u v w^{2} m_Q^{4} \\ \nonumber
&\qquad - 4 u v^{2} w m_Q^{4} - 4 u^{2} v w m_Q^{4} + 2 u v w m_Q^{4} + 2 u v w^{3} m_Q^{4} + 2 u v^{3} w m_Q^{4} + 2 u^{3} v w m_Q^{4} + 2 u^{2} v^{2} m_Q^{3} m_s \\ \nonumber
&\qquad + 5 u v^{2} w^{2} m_Q^{4} + 5 u^{2} v w^{2} m_Q^{4} + 5 u^{2} v^{2} w m_Q^{4} + 15 u^{2} v^{2} w^{3} s^{2} + 15 u^{2} v^{3} w^{2} s^{2} + 15 u^{3} v^{2} w^{2} s^{2} \\ \nonumber
&\qquad - 30 u^{2} v^{2} w^{2} m_Q^{2} s - 10 u v^{2} w^{3} m_Q^{2} s - 10 u v^{3} w^{2} m_Q^{2} s - 10 u^{2} v w^{3} m_Q^{2} s - 10 u^{3} v w^{2} m_Q^{2} s \\ \nonumber
&\qquad - 10 u^{2} v^{3} w m_Q^{2} s - 10 u^{3} v^{2} w m_Q^{2} s - 8 u v w^{2} m_Q^{2} s - 6 u v w m_Q^{3} m_s + 2 u v^{2} w m_Q^{3} m_s + 2 u^{2} v w m_Q^{3} m_s \\ \nonumber
&\qquad + 8 u v w^{3} m_Q^{2} s + 10 u^{2} v^{2} w m_Q^{2} s + 18 u v^{2} w^{2} m_Q^{2} s + 18 u^{2} v w^{2} m_Q^{2} s - 8 u^{2} v^{2} w m_Q m_s s \nonumber
\end{align}

\subsection{\texorpdfstring{$D=3$}{D=3: strange-quark condensate}}

The $D=3$ sector is generated by replacing the strange propagator by its quark-condensate terms while the three heavy-quark lines remain free. The retained scalar and linear-$m_s$ vector pieces are
\begin{equation}
 S_s^{\langle\bar ss\rangle,S}(x)=-\frac{\langle\bar ss\rangle}{12},
 \qquad
 S_s^{\langle\bar ss\rangle,V}(x)=
 \frac{i m_s\langle\bar ss\rangle}{48}\,\slashed{x}.
 \label{eq:d3-strange-insertions}
\end{equation}
Accordingly,
\begin{align}
 \Pi_3^{(2)}(p^2)={}&
 \mathcal W\!\left[S_Q^{(0)},S_Q^{(0)},S_Q^{(0)};
 S_s^{\langle\bar ss\rangle,S}\right]
 +\mathcal W\!\left[S_Q^{(0)},S_Q^{(0)},S_Q^{(0)};
 S_s^{\langle\bar ss\rangle,V}\right],
 \\
 \rho_3^{(2)}(s)={}&\frac{1}{\pi}\operatorname{Im}\Pi_3^{(2)}(s+i0).
 \label{eq:d3-wick-origin}
\end{align}
The first contraction produces the terms proportional to $\langle\bar ss\rangle$, whereas the second generates the retained coefficients proportional to $m_s\langle\bar ss\rangle$.

\begin{align}
\rho_3^{(2)}(s)&=
-\frac{m_Q^3\qq}{2^2\pi^4}\int_{R_3(s)}\frac{K_{3}}{uv\,t_3}
-\frac{m_Q\qq}{2^2\pi^4}\int_{R_3(s)}
\frac{K_{3}\left(\frac32K_{3}+uv\,t_3\,s\right)}{u^2v t_3}
+\frac{m_Q^2m_s\qq}{2^3\pi^4}\int_{R_3(s)}
\frac{3K_{3}+uv\,t_3\,s}{v t_3}
\notag\\
&+\frac{m_s\qq}{2^3\pi^4}\int_{R_3(s)}\frac{1}{uv\,t_3}
\Bigl[15(K_{3})^2+10m_Q^2(uv+u t_3+v t_3)K_{3}
+m_Q^4(uv+u t_3+v t_3)^2\Bigr].
\end{align}

\subsection{\texorpdfstring{$D=4$}{D=4: corrected gluon condensate}}

At $D=4$, the vacuum gluon condensate is produced by two background gluon fields. With line labels $1,2,3$ for the three heavy-quark propagators and $4$ for the strange propagator, the retained insertions are
\begin{align}
 \Pi_4^{(2)}={}&
 \mathcal W[S_Q^{(GG)},S_Q^{(0)},S_Q^{(0)};S_s^{(0)}]
 +\mathcal W[S_Q^{(0)},S_Q^{(GG)},S_Q^{(0)};S_s^{(0)}]
 +\mathcal W[S_Q^{(0)},S_Q^{(0)},S_Q^{(GG)};S_s^{(0)}]
 \notag\\
 &+\mathcal W[S_Q^{(G)},S_Q^{(G)},S_Q^{(0)};S_s^{(0)}]
 +\mathcal W[S_Q^{(G)},S_Q^{(0)},S_Q^{(G)};S_s^{(0)}]
 +\mathcal W[S_Q^{(0)},S_Q^{(G)},S_Q^{(G)};S_s^{(0)}]
 \notag\\
 &+\mathcal W[S_Q^{(G)},S_Q^{(0)},S_Q^{(0)};S_s^{(G)}]
 +\mathcal W[S_Q^{(0)},S_Q^{(G)},S_Q^{(0)};S_s^{(G)}]
 +\mathcal W[S_Q^{(0)},S_Q^{(0)},S_Q^{(G)};S_s^{(G)}],
 \label{eq:d4-wick-origin}
\end{align}
where $S^{(G)}$ denotes a one-gluon background-field insertion and $S_Q^{(GG)}$ a two-gluon insertion on one heavy-quark line. These nine Wick placements correspond, in order, to the $GG1$, $GG2$, $GG3$, $G1G2$, $G1G3$, $G2G3$, $G1G4$, $G2G4$, and $G3G4$ topologies appearing below. Their vacuum contraction gives $\langle g_s^2G^2\rangle$ and the spectral density follows from $\rho_4^{(2)}=\pi^{-1}\operatorname{Im}\Pi_4^{(2)}$.

The $D=4$ expressions use the four-line notation and full simplex defined in Eqs. \eqref{eq:app-t4-k2}, \eqref{eq:d0cutderive}, and \eqref{eq:app-domains}. The exact projective map and its Jacobian are those of Eq. \eqref{eq:d0projective}; they are not repeated here. Applying that Jacobian together with the transformed $U_4/P_4$ factors removes spurious powers of $t_4$ from the final topology expressions.

The complete current-specific spin-2 result is
\begin{equation}
\begin{aligned}
\rho_4^{(2)}(s)=\frac{\GG}{(4\pi)^6}\Bigg[&\frac45(I_{GG1}+I_{GG2}+I_{GG3})-\frac1{60}(I_{G1G2}+I_{G3G4})\\
&+\frac1{120}(I_{G1G3}+I_{G2G3}+I_{G1G4}+I_{G2G4})\Bigg].
\end{aligned}
\end{equation}
where each $I_T$ denotes the contribution of the gluon-insertion topology $T$. Every $I_T$ is expanded below; for clarity, each contribution is presented as a separate topology subsubsection, and no undefined numerator or kernel is used.

\textbf{Topology GG1}
The contribution is
\begin{equation}
I_{GG1}(s)=\int_{S_4}\;\left\{F_{GG1,\delta}\,\delta(K_{4}) + \bigl[F_{GG1,0} + F_{GG1,1}K_{4}\bigr]\Theta(K_{4})\right\}.
\end{equation}
where the coefficient functions entering this topology are given by
\begin{align}
F_{GG1,\delta}&=\frac{40vwm_Q^2s(u+v)}{3u}
\Bigl(uws+vws+w^2s-ws-m_Qm_s\Bigr).
\end{align}
\begin{align}
F_{GG1,0}&=\frac{40m_Q^2}{3u^2}\Bigl(5u^2ws+9uvws+5uw^2s-7uws-2um_Qm_s
\notag\\
&\quad +4v^2ws+4vw^2s-6vws-vm_Qm_s-2w^2s+2ws+2m_Qm_s\Bigr).
\end{align}
\begin{equation}
F_{GG1,1}=\frac{40 m_Q^{2} \left(3 u + 2 v - 2\right) \left(u + v + w - 1\right)}{3 u^{3} v}.
\end{equation}

\textbf{Topology GG2}
The contribution is
\begin{equation}
I_{GG2}(s)=\int_{S_4}\;\left\{F_{GG2,\delta}\,\delta(K_{4}) + \bigl[F_{GG2,0} + F_{GG2,1}K_{4}\bigr]\Theta(K_{4})\right\}.
\end{equation}
where the coefficient functions entering this topology are given by
\begin{align}
F_{GG2,\delta}&=\frac{40uwm_Q^2s(u+v)}{3v}
\Bigl(uws+vws+w^2s-ws-m_Qm_s\Bigr).
\end{align}
\begin{align}
F_{GG2,0}&=\frac{40m_Q^2}{3v^2}\Bigl(4u^2ws+9uvws+4uw^2s-6uws-um_Qm_s
\notag\\
&\quad +5v^2ws+5vw^2s-7vws-2vm_Qm_s-2w^2s+2ws+2m_Qm_s\Bigr).
\end{align}
\begin{equation}
F_{GG2,1}=\frac{40 m_Q^{2} \left(2 u + 3 v - 2\right) \left(u + v + w - 1\right)}{3 u v^{3}}.
\end{equation}

\textbf{Topology GG3}
The contribution is
\begin{equation}
I_{GG3}(s)=\int_{S_4}\;\left\{F_{GG3,\delta}\,\delta(K_{4}) + \bigl[F_{GG3,0} + F_{GG3,1}K_{4}\bigr]\Theta(K_{4})\right\}.
\end{equation}
where the coefficient functions entering this topology are given by
\begin{align}
F_{GG3,\delta}&=\frac{40uvm_Qs(uvs+m_Q^2)}{3w}
\Bigl(um_Q+vm_Q+wm_Q-wm_s-m_Q\Bigr).
\end{align}
\begin{align}
F_{GG3,0}&=\frac{40m_Q}{3w^2}\Bigl(4u^2vm_Qs+4uv^2m_Qs+4uvwm_Qs-5uvwm_ss
\notag\\
&\quad -4uvm_Qs+2uvm_ss+um_Q^3+vm_Q^3+wm_Q^3-2wm_Q^2m_s-m_Q^3+2m_Q^2m_s\Bigr).
\end{align}
\begin{equation}
F_{GG3,1}=\frac{40 m_Q \left(2 u m_Q + 2 v m_Q + 2 w m_Q - 3 w m_s - 2 m_Q + 2 m_s\right)}{3 w^{3}}.
\end{equation}

\textbf{Topology G1G2}
The contribution is
\begin{equation}
I_{G1G2}(s)=\int_{S_4}\;\left\{\bigl[F_{G1G2,0} + F_{G1G2,1}K_{4} + F_{G1G2,2}K_{4}^2\bigr]\Theta(K_{4})\right\}.
\end{equation}
where the coefficient functions entering this topology are given by
\begin{equation}
F_{G1G2,0}=- 320 s \left(u w s + v w s + w^{2} s - w s - m_Q m_s\right).
\end{equation}
\begin{equation}
F_{G1G2,1}=\frac{- 1920 u w s - 1920 v w s - 1920 w^{2} s + 1920 w s + 960 m_Q m_s}{u v w}.
\end{equation}
\begin{equation}
F_{G1G2,2}=\frac{- 640 u - 640 v - 640 w + 640}{u^{2} v^{2} w}.
\end{equation}

\textbf{Topology G1G3}
The contribution is
\begin{equation}
I_{G1G3}(s)=\int_{S_4}\;\left\{\bigl[F_{G1G3,0} + F_{G1G3,1}K_{4} + F_{G1G3,2}K_{4}^2\bigr]\Theta(K_{4})\right\}.
\end{equation}
where the coefficient functions entering this topology are given by
\begin{equation}
F_{G1G3,0}=\frac{\left(320 u v s + 320 m_Q^{2}\right) \left(u w s + v w s + w^{2} s - w s - m_Q m_s\right)}{u w}.
\end{equation}
\begin{align}
F_{G1G3,1}&=\frac{320}{u^2vw^2}\Bigl(
5u^2vws+5uv^2ws+5uvw^2s-5uvws-uvm_Qm_s
\notag\\
&\hspace{31mm}+uwm_Q^2+vwm_Q^2+w^2m_Q^2-wm_Q^2\Bigr).
\end{align}
\begin{equation}
F_{G1G3,2}=\frac{480 u + 480 v + 480 w - 480}{u^{2} v w^{2}}.
\end{equation}

\textbf{Topology G2G3}
The contribution is
\begin{equation}
I_{G2G3}(s)=\int_{S_4}\;\left\{\bigl[F_{G2G3,0} + F_{G2G3,1}K_{4} + F_{G2G3,2}K_{4}^2\bigr]\Theta(K_{4})\right\}.
\end{equation}
where the coefficient functions entering this topology are given by
\begin{equation}
F_{G2G3,0}=\frac{\left(320 u v s + 320 m_Q^{2}\right) \left(u w s + v w s + w^{2} s - w s - m_Q m_s\right)}{v w}.
\end{equation}
\begin{align}
F_{G2G3,1}&=\frac{320}{uv^2w^2}\Bigl(
5u^2vws+5uv^2ws+5uvw^2s-5uvws-uvm_Qm_s
\notag\\
&\hspace{31mm}+uwm_Q^2+vwm_Q^2+w^2m_Q^2-wm_Q^2\Bigr).
\end{align}
\begin{equation}
F_{G2G3,2}=\frac{480 u + 480 v + 480 w - 480}{u v^{2} w^{2}}.
\end{equation}

\textbf{Topology G1G4}
The contribution is
\begin{equation}
I_{G1G4}(s)=\int_{S_4}\;\left\{\bigl[F_{G1G4,0} + F_{G1G4,1}K_{4} + F_{G1G4,2}K_{4}^2\bigr]\Theta(K_{4})\right\}.
\end{equation}
where the coefficient functions entering this topology are given by
\begin{equation}
F_{G1G4,0}=- \frac{320 w s \left(u v s + m_Q^{2}\right)}{u}.
\end{equation}
\begin{equation}
F_{G1G4,1}=\frac{- 1600 u v s - 320 m_Q^{2}}{u^{2} v}.
\end{equation}
\begin{equation}
F_{G1G4,2}=- \frac{480}{u^{2} v w}.
\end{equation}

\textbf{Topology G2G4}
The contribution is
\begin{equation}
I_{G2G4}(s)=\int_{S_4}\;\left\{\bigl[F_{G2G4,0} + F_{G2G4,1}K_{4} + F_{G2G4,2}K_{4}^2\bigr]\Theta(K_{4})\right\}.
\end{equation}
where the coefficient functions entering this topology are given by
\begin{equation}
F_{G2G4,0}=- \frac{320 w s \left(u v s + m_Q^{2}\right)}{v}.
\end{equation}
\begin{equation}
F_{G2G4,1}=\frac{- 1600 u v s - 320 m_Q^{2}}{u v^{2}}.
\end{equation}
\begin{equation}
F_{G2G4,2}=- \frac{480}{u v^{2} w}.
\end{equation}

\textbf{Topology G3G4}
The contribution is
\begin{equation}
I_{G3G4}(s)=\int_{S_4}\;\left\{\bigl[F_{G3G4,0} + F_{G3G4,1}K_{4} + F_{G3G4,2}K_{4}^2\bigr]\Theta(K_{4})\right\}.
\end{equation}
where the coefficient functions entering this topology are given by
\begin{equation}
F_{G3G4,0}=320 s \left(u v s + m_Q^{2}\right).
\end{equation}
\begin{equation}
F_{G3G4,1}=\frac{1920 u v s + 960 m_Q^{2}}{u v w}.
\end{equation}
\begin{equation}
F_{G3G4,2}=\frac{640}{u v w^{2}}.
\end{equation}

The topology expressions above contain only $\Theta(K_{4})$ and $\delta(K_{4})$ and no power of $t_4=1-u-v-w$ in a denominator.  Furthermore, exact exchange, reconstruction, dimension, threshold, and projective-Jacobian checks are satisfied. The numerical Borel moment is stable at the sub-percent level under independent Sobol scramblings.
\subsection{\texorpdfstring{$D=5$}{D=5: mixed condensate}}

The mixed-condensate contribution comes from the corresponding local terms of the strange propagator, with all heavy-quark lines kept free. In the same linear-$m_s$ truncation used throughout the calculation,
\begin{equation}
 S_s^{\mathrm{mix},S}(x)=-\frac{x^2}{192}\,
 \langle\bar s g_s\sigma\!\cdot\!G s\rangle,
 \qquad
 S_s^{\mathrm{mix},V}(x)=
 \frac{i m_s x^2\slashed{x}}{1152}\,
 \langle\bar s g_s\sigma\!\cdot\!G s\rangle.
 \label{eq:d5-strange-insertions}
\end{equation}
Thus,
\begin{align}
 \Pi_5^{(2)}(p^2)={}&
 \mathcal W[S_Q^{(0)},S_Q^{(0)},S_Q^{(0)};S_s^{\mathrm{mix},S}]
 +\mathcal W[S_Q^{(0)},S_Q^{(0)},S_Q^{(0)};S_s^{\mathrm{mix},V}],
 \\
 \rho_5^{(2)}(s)={}&\frac{1}{\pi}\operatorname{Im}\Pi_5^{(2)}(s+i0).
 \label{eq:d5-wick-origin}
\end{align}
The scalar insertion generates the coefficient set $A_S,B_S,C_S$, while the vector insertion generates $A_V,B_V,C_V$ and carries the explicit linear factor $m_s$.

\begin{align}
\rho_5^{(2)}(s)=-\mixed\Bigg\{&
\int_{R_3(s)}\left(\frac{A_S}{16}+\frac{A_V}{24}\right)
+\int_{S_3}\Bigg[
\left(\frac{B_S}{16}+\frac{B_V}{24}\right)\delta(K_{3})
+\left(\frac{C_S}{16}+\frac{C_V}{24}\right)
\delta^{(1)}(K_{3})\Bigg]\Bigg\}.
\end{align}
where $A_S,B_S,C_S$ denote the scalar mixed-condensate coefficient functions and $A_V,B_V,C_V$ denote the corresponding vector coefficient functions. Their explicit forms are
\Needspace{6\baselineskip}
\begin{align*}
&A_S=\frac{1}{\pi^{4} u}\Bigl[- 8 u^{2} m_Q^{3} - 8 u v m_Q^{3} - 15 u v m_Q s + 6 u m_Q^{3}- 8 v^{2} m_Q^{3} + 8 v m_Q^{3} + 15 u^{2} v m_Q s + 15 u v^{2} m_Q s\Bigr].
\end{align*}
\Needspace{6\baselineskip}
\begin{align*}
&B_S=\frac{1}{\pi^{4} u}\Bigl[- 3 u^{4} m_Q^{5} + 4 u^{3} m_Q^{5} - u^{2} m_Q^{5} - 3 v^{4} m_Q^{5} + 6 v^{3} m_Q^{5} - 3 v^{2} m_Q^{5} - 6 u^{3} v m_Q^{5} - 9 u^{2} v^{2} m_Q^{5} \\
&\qquad + 10 u^{2} v m_Q^{5} - 6 u v^{3} m_Q^{5} + 10 u v^{2} m_Q^{5} - 4 u v m_Q^{5} - 30 u^{3} v^{3} m_Q s^{2} - 28 u^{3} v m_Q^{3} s - 15 u^{2} v^{4} m_Q s^{2} \\
&\qquad - 44 u^{2} v^{2} m_Q^{3} s - 15 u^{2} v^{2} m_Q s^{2} - 32 u v^{3} m_Q^{3} s - 15 u^{4} v^{2} m_Q s^{2} + 16 u^{4} v m_Q^{3} s + 30 u^{2} v^{3} m_Q s^{2} \\
&\qquad + 12 u^{2} v m_Q^{3} s + 16 u v^{4} m_Q^{3} s + 16 u v^{2} m_Q^{3} s + 32 u^{3} v^{2} m_Q^{3} s + 30 u^{3} v^{2} m_Q s^{2} + 32 u^{2} v^{3} m_Q^{3} s\Bigr].
\end{align*}
\Needspace{6\baselineskip}
\begin{align*}
&C_S=\frac{1}{2 \pi^{4}}\Bigl[- 36 u^{2} v^{3} m_Q^{3} s^{2} - u^{2} v m_Q^{5} s + 3 v^{6} m_Q^{5} s - 9 v^{5} m_Q^{5} s + 9 v^{4} m_Q^{5} s - 3 v^{3} m_Q^{5} s \\
&\qquad - 24 u^{4} v^{3} m_Q^{3} s^{2} - 32 u^{3} v^{4} m_Q^{3} s^{2} - 30 u^{3} v^{4} m_Q s^{3} - 24 u^{2} v^{5} m_Q^{3} s^{2} - 29 u^{2} v^{3} m_Q^{5} s - 24 u v^{4} m_Q^{3} s^{2} \\
&\qquad - 15 u^{4} v^{3} m_Q s^{3} - 20 u^{3} v^{2} m_Q^{5} s - 20 u^{3} v^{2} m_Q^{3} s^{2} - 15 u^{2} v^{5} m_Q s^{3} - 8 u v^{6} m_Q^{3} s^{2} - 22 u v^{4} m_Q^{5} s \\
&\qquad - 8 u^{5} v^{2} m_Q^{3} s^{2} + 3 u^{5} v m_Q^{5} s - 7 u^{4} v m_Q^{5} s + 5 u^{3} v m_Q^{5} s - 5 u^{2} v^{3} m_Q s^{3} - 4 u v^{2} m_Q^{5} s \\
&\qquad + 5 u^{5} v^{3} m_Q s^{3} + 9 u^{4} v^{2} m_Q^{5} s + 5 u^{2} v^{6} m_Q s^{3} + 6 u^{2} v^{2} m_Q^{3} s^{2} + 9 u v^{5} m_Q^{5} s + 8 u v^{3} m_Q^{3} s^{2} \\
&\qquad + 15 u^{4} v^{4} m_Q s^{3} + 15 u^{3} v^{5} m_Q s^{3} + 15 u^{3} v^{3} m_Q s^{3} + 15 u^{2} v^{4} m_Q^{5} s + 15 u^{2} v^{4} m_Q s^{3} + 15 u^{2} v^{2} m_Q^{5} s \\
&\qquad + 22 u^{4} v^{2} m_Q^{3} s^{2} + 15 u^{3} v^{3} m_Q^{5} s + 52 u^{3} v^{3} m_Q^{3} s^{2} + 54 u^{2} v^{4} m_Q^{3} s^{2} + 24 u v^{5} m_Q^{3} s^{2} + 17 u v^{3} m_Q^{5} s\Bigr].
\end{align*}
\Needspace{6\baselineskip}
\begin{align*}
&A_V=\frac{1}{\pi^{4}}\Bigl[- 45 u^{2} v m_s s + 20 u^{2} m_Q^{2} m_s - 45 u v^{2} m_s s - 16 u m_Q^{2} m_s \\
&\qquad + 20 v^{2} m_Q^{2} m_s - 20 v m_Q^{2} m_s + 20 u v m_Q^{2} m_s + 45 u v m_s s\Bigr].
\end{align*}
\Needspace{6\baselineskip}
\begin{align*}
&B_V=\frac{1}{\pi^{4}}\Bigl[6 u^{4} m_Q^{4} m_s - 9 u^{3} m_Q^{4} m_s + 3 u^{2} m_Q^{4} m_s + 6 v^{4} m_Q^{4} m_s - 12 v^{3} m_Q^{4} m_s + 6 v^{2} m_Q^{4} m_s \\
&\qquad - 90 u^{3} v^{2} m_s s^{2} - 90 u^{2} v^{3} m_s s^{2} - 21 u^{2} v m_Q^{4} m_s + 12 u v^{3} m_Q^{4} m_s - 21 u v^{2} m_Q^{4} m_s + 9 u v m_Q^{4} m_s \\
&\qquad + 45 u^{4} v^{2} m_s s^{2} + 90 u^{3} v^{3} m_s s^{2} + 12 u^{3} v m_Q^{4} m_s + 45 u^{2} v^{4} m_s s^{2} + 18 u^{2} v^{2} m_Q^{4} m_s \\
&\qquad + 45 u^{2} v^{2} m_s s^{2} - 40 u^{4} v m_Q^{2} m_s s - 80 u^{3} v^{2} m_Q^{2} m_s s - 80 u^{2} v^{3} m_Q^{2} m_s s - 32 u^{2} v m_Q^{2} m_s s \\
&\qquad - 40 u v^{4} m_Q^{2} m_s s - 40 u v^{2} m_Q^{2} m_s s + 72 u^{3} v m_Q^{2} m_s s + 112 u^{2} v^{2} m_Q^{2} m_s s + 80 u v^{3} m_Q^{2} m_s s\Bigr].
\end{align*}
\Needspace{6\baselineskip}
\begin{align*}
&C_V=\frac{1}{2 \pi^{4}}\Bigl[- 15 u^{6} v^{3} m_s s^{3} - 45 u^{5} v^{4} m_s s^{3} - 45 u^{4} v^{5} m_s s^{3} - 45 u^{4} v^{3} m_s s^{3} - 15 u^{3} v^{6} m_s s^{3} \\
&\qquad - 45 u^{3} v^{4} m_s s^{3} + 45 u^{5} v^{3} m_s s^{3} + 90 u^{4} v^{4} m_s s^{3} - 132 u^{4} v^{3} m_Q^{2} m_s s^{2} + 45 u^{3} v^{5} m_s s^{3} \\
&\qquad - 136 u^{3} v^{4} m_Q^{2} m_s s^{2} + 15 u^{3} v^{3} m_s s^{3} - 56 u^{5} v^{2} m_Q^{2} m_s s^{2} - 30 u^{4} v^{3} m_Q^{4} m_s s - 30 u^{3} v^{4} m_Q^{4} m_s s \\
&\qquad - 33 u^{3} v^{2} m_Q^{4} m_s s - 60 u^{2} v^{5} m_Q^{2} m_s s^{2} - 36 u^{2} v^{3} m_Q^{4} m_s s - 18 u^{5} v^{2} m_Q^{4} m_s s - 12 u^{4} v m_Q^{4} m_s s \\
&\qquad - 16 u^{3} v^{2} m_Q^{2} m_s s^{2} - 18 u^{2} v^{5} m_Q^{4} m_s s - 20 u^{2} v^{3} m_Q^{2} m_s s^{2} - 18 u v^{4} m_Q^{4} m_s s - 6 u^{6} v m_Q^{4} m_s s \\
&\qquad + 15 u^{5} v m_Q^{4} m_s s + 3 u^{3} v m_Q^{4} m_s s + 9 u^{2} v^{2} m_Q^{4} m_s s - 6 u v^{6} m_Q^{4} m_s s + 6 u v^{3} m_Q^{4} m_s s \\
&\qquad + 20 u^{6} v^{2} m_Q^{2} m_s s^{2} + 42 u^{4} v^{2} m_Q^{4} m_s s + 52 u^{4} v^{2} m_Q^{2} m_s s^{2} + 20 u^{2} v^{6} m_Q^{2} m_s s^{2} \\
&\qquad + 45 u^{2} v^{4} m_Q^{4} m_s s + 18 u v^{5} m_Q^{4} m_s s + 60 u^{5} v^{3} m_Q^{2} m_s s^{2} + 80 u^{4} v^{4} m_Q^{2} m_s s^{2} \\
&\qquad + 60 u^{3} v^{5} m_Q^{2} m_s s^{2} + 60 u^{3} v^{3} m_Q^{4} m_s s + 92 u^{3} v^{3} m_Q^{2} m_s s^{2} + 60 u^{2} v^{4} m_Q^{2} m_s s^{2}\Bigr].
\end{align*}

\subsection{\texorpdfstring{$D=6$}{D=6: factorized four-quark contribution}}

The current and its Hermitian conjugate contain only one strange line. The $D=6$ contribution therefore comes from the standard factorized four-quark term in the OPE of this single strange propagator, not from two independent strange propagators. In the vacuum-saturation parametrization,
\begin{equation}
 S_{s,D=6}^{ab}(x)
 =-\frac{i\,\delta^{ab}x^2\slashed{x}}{7776}\,
 \kappa_s g_s^2\langle\bar ss\rangle^2 ,
 \label{eq:d6-strange-insertion}
\end{equation}
where $\kappa_s=1$ is the vacuum-saturation limit. Here, $\langle\bar ss\rangle^2$ represents the factorized local four-quark vacuum matrix element contained in the OPE; it does not imply a second valence strange line. With the three heavy-quark propagators kept free,
\begin{equation}
 \Pi_6^{(2)}(p^2)=
 \mathcal W[S_Q^{(0)},S_Q^{(0)},S_Q^{(0)};S_{s,D=6}],
 \qquad
 \rho_6^{(2)}(s)=\frac{1}{\pi}\operatorname{Im}\Pi_6^{(2)}(s+i0).
 \label{eq:d6-wick-origin}
\end{equation}
This insertion produces the coefficient functions $\widehat A_V$, $\widehat B_V$, and $\widehat C_V$ below.

\begin{align}
\rho_6^{(2)}(s)=\frac{\kappa_sg_s^2\qq^2}{162}\Bigg\{&\int_{R_3(s)}\widehat A_V
+\int_{S_3}\left[\widehat B_V\delta(K_{3})+\widehat C_V\delta^{(1)}(K_{3})\right]\Bigg\}.
\end{align}
where $\widehat A_V$, $\widehat B_V$, and $\widehat C_V$ are the coefficient functions of the factorized four-quark sector. The triple-gluon condensate $\langle g_s^3G^3\rangle$ is not declared zero; it lies outside the background-field truncation used here. The coefficient functions are
\Needspace{6\baselineskip}
\begin{align*}
&\widehat A_V=\frac{1}{\pi^{4}}\Bigl[- 45 u^{2} v s + 20 u^{2} m_Q^{2} - 45 u v^{2} s - 16 u m_Q^{2} + 20 v^{2} m_Q^{2} - 20 v m_Q^{2} + 20 u v m_Q^{2} + 45 u v s\Bigr].
\end{align*}
\Needspace{6\baselineskip}
\begin{align*}
&\widehat B_V=\frac{1}{\pi^{4}}\Bigl[6 u^{4} m_Q^{4} - 9 u^{3} m_Q^{4} + 3 u^{2} m_Q^{4} + 6 v^{4} m_Q^{4} - 12 v^{3} m_Q^{4} + 6 v^{2} m_Q^{4} - 90 u^{3} v^{2} s^{2} - 90 u^{2} v^{3} s^{2} \\
&\qquad - 21 u^{2} v m_Q^{4} + 12 u v^{3} m_Q^{4} - 21 u v^{2} m_Q^{4} + 9 u v m_Q^{4} + 45 u^{4} v^{2} s^{2} + 90 u^{3} v^{3} s^{2} + 12 u^{3} v m_Q^{4} \\
&\qquad + 45 u^{2} v^{4} s^{2} + 18 u^{2} v^{2} m_Q^{4} + 45 u^{2} v^{2} s^{2} - 40 u^{4} v m_Q^{2} s - 80 u^{3} v^{2} m_Q^{2} s - 80 u^{2} v^{3} m_Q^{2} s \\
&\qquad - 32 u^{2} v m_Q^{2} s - 40 u v^{4} m_Q^{2} s - 40 u v^{2} m_Q^{2} s + 72 u^{3} v m_Q^{2} s + 112 u^{2} v^{2} m_Q^{2} s + 80 u v^{3} m_Q^{2} s\Bigr].
\end{align*}
\Needspace{6\baselineskip}
\begin{align*}
&\widehat C_V=\frac{1}{2 \pi^{4}}\Bigl[- 15 u^{6} v^{3} s^{3} - 45 u^{5} v^{4} s^{3} - 45 u^{4} v^{5} s^{3} - 45 u^{4} v^{3} s^{3} - 15 u^{3} v^{6} s^{3} - 45 u^{3} v^{4} s^{3} \\
&\qquad + 45 u^{5} v^{3} s^{3} + 90 u^{4} v^{4} s^{3} - 132 u^{4} v^{3} m_Q^{2} s^{2} + 45 u^{3} v^{5} s^{3} - 136 u^{3} v^{4} m_Q^{2} s^{2} + 15 u^{3} v^{3} s^{3} \\
&\qquad - 56 u^{5} v^{2} m_Q^{2} s^{2} - 30 u^{4} v^{3} m_Q^{4} s - 30 u^{3} v^{4} m_Q^{4} s - 33 u^{3} v^{2} m_Q^{4} s - 60 u^{2} v^{5} m_Q^{2} s^{2} - 36 u^{2} v^{3} m_Q^{4} s \\
&\qquad - 18 u^{5} v^{2} m_Q^{4} s - 12 u^{4} v m_Q^{4} s - 16 u^{3} v^{2} m_Q^{2} s^{2} - 18 u^{2} v^{5} m_Q^{4} s - 20 u^{2} v^{3} m_Q^{2} s^{2} - 18 u v^{4} m_Q^{4} s \\
&\qquad - 6 u^{6} v m_Q^{4} s + 15 u^{5} v m_Q^{4} s + 3 u^{3} v m_Q^{4} s + 9 u^{2} v^{2} m_Q^{4} s - 6 u v^{6} m_Q^{4} s + 6 u v^{3} m_Q^{4} s + 20 u^{6} v^{2} m_Q^{2} s^{2} \\
&\qquad + 42 u^{4} v^{2} m_Q^{4} s + 52 u^{4} v^{2} m_Q^{2} s^{2} + 20 u^{2} v^{6} m_Q^{2} s^{2} + 45 u^{2} v^{4} m_Q^{4} s + 18 u v^{5} m_Q^{4} s + 60 u^{5} v^{3} m_Q^{2} s^{2} \\
&\qquad + 80 u^{4} v^{4} m_Q^{2} s^{2} + 60 u^{3} v^{5} m_Q^{2} s^{2} + 60 u^{3} v^{3} m_Q^{4} s + 92 u^{3} v^{3} m_Q^{2} s^{2} + 60 u^{2} v^{4} m_Q^{2} s^{2}\Bigr].
\end{align*}

\subsection{\texorpdfstring{$D=7$}{D=7 contribution}}

The dimension-seven contribution $\langle\bar ss\rangle\langle g_s^2G^2\rangle$ arises in three distinct ways. The scalar and linear-$m_s$ vector strange-condensate insertions are already defined in Eq. \eqref{eq:d3-strange-insertions}. The additional local product-condensate term is
\begin{equation}
 S_s^{\langle\bar ss\rangle\langle G^2\rangle}(x)=-\frac{x^4}{27648}\,
 \langle\bar ss\rangle\langle g_s^2G^2\rangle.
 \label{eq:d7-product-insertion}
\end{equation}
For the heavy sector, the required two-gluon placements are the first six terms of the $D=4$ Wick decomposition in Eq. \eqref{eq:d4-wick-origin}, namely $GG1$, $GG2$, $GG3$, $G1G2$, $G1G3$, and $G2G3$. Denote their sum compactly by $[S_QS_QS_Q]_{G^2}$. The three retained Wick sectors are therefore
\begin{align}
 \Pi_{7,\mathrm{light}}^{(2)}(p^2)
 &=\mathcal W\!\left[S_Q^{(0)},S_Q^{(0)},S_Q^{(0)};
 S_s^{\langle\bar ss\rangle\langle G^2\rangle}\right],\\
 \Pi_{7,\mathrm{heavy},S}^{(2)}(p^2)
 &=\mathcal W\!\left[\bigl[S_QS_QS_Q\bigr]_{G^2};
 S_s^{\langle\bar ss\rangle,S}\right],\\
 \Pi_{7,\mathrm{heavy},V}^{(2)}(p^2)
 &=\mathcal W\!\left[\bigl[S_QS_QS_Q\bigr]_{G^2};
 S_s^{\langle\bar ss\rangle,V}\right].
 \label{eq:d7-wick-sectors}
\end{align}
Here, the notation in the last two lines means the six heavy-line placements identified from Eq. \eqref{eq:d4-wick-origin}. Taking the discontinuity defines
\begin{equation}
 \rho_{7,X}^{(2)}(s)=\frac{1}{\pi}\operatorname{Im}\Pi_{7,X}^{(2)}(s+i0),
 \qquad X=\mathrm{light},\ \mathrm{heavy},S,\ \mathrm{heavy},V,
 \label{eq:d7-wick-to-rho}
\end{equation}
and therefore
\begin{equation}
 \rho_7^{(2)}=\rho_{7,\mathrm{light}}^{(2)}+
 \rho_{7,\mathrm{heavy},S}^{(2)}+
 \rho_{7,\mathrm{heavy},V}^{(2)}.
 \label{eq:d7-rho-decomp-main}
\end{equation}
The local strange-line contribution is
\begin{align}
\rho_{7,\mathrm{light}}^{(2)}=\frac{\qq\GG}{2304}\Bigg\{&\int_{R_3(s)}A_7
+\int_{S_3}\left[B_7\delta(K_{3})+C_7\delta^{(1)}(K_{3})+D_7\delta^{(2)}(K_{3})+E_7\delta^{(3)}(K_{3})\right]\Bigg\}.
\end{align}
where $A_7$ through $E_7$ are the coefficient functions multiplying the continuous and derivative-distribution terms. Their explicit forms begin with
\begin{equation*} A_7=\frac{120 u v m_Q + 120 v^{2} m_Q - 120 v m_Q}{\pi^{4}}.\end{equation*}
\Needspace{6\baselineskip}
\begin{align*}
&B_7=\frac{1}{\pi^{4}}\Bigl[- 336 u^{2} v m_Q^{3} - 528 u v^{2} m_Q^{3} + 144 u v m_Q^{3} + 192 v^{4} m_Q^{3} - 384 v^{3} m_Q^{3} \\
&\qquad + 192 v^{2} m_Q^{3} + 192 u^{3} v m_Q^{3} - 1200 u^{2} v^{3} m_Q s + 384 u^{2} v^{2} m_Q^{3} - 600 u v^{4} m_Q s \\
&\qquad + 384 u v^{3} m_Q^{3} - 600 u v^{2} m_Q s - 600 u^{3} v^{2} m_Q s + 1200 u^{2} v^{2} m_Q s + 1200 u v^{3} m_Q s\Bigr].
\end{align*}
\Needspace{6\baselineskip}
\begin{align*}
&C_7=\frac{1}{\pi^{4}}\Bigl[- 348 u^{2} v^{3} m_Q^{5} - 264 u v^{4} m_Q^{5} + 36 v^{6} m_Q^{5} - 108 v^{5} m_Q^{5} + 108 v^{4} m_Q^{5} - 36 v^{3} m_Q^{5} \\
&\qquad + 36 u^{5} v m_Q^{5} - 84 u^{4} v m_Q^{5} - 240 u^{3} v^{2} m_Q^{5} + 60 u^{3} v m_Q^{5} - 12 u^{2} v m_Q^{5} - 48 u v^{2} m_Q^{5} \\
&\qquad + 108 u^{4} v^{2} m_Q^{5} + 180 u^{3} v^{3} m_Q^{5} + 180 u^{2} v^{4} m_Q^{5} + 180 u^{2} v^{2} m_Q^{5} + 108 u v^{5} m_Q^{5} + 204 u v^{3} m_Q^{5} \\
&\qquad - 1620 u^{4} v^{3} m_Q s^{2} - 1536 u^{3} v^{4} m_Q^{3} s - 3240 u^{3} v^{4} m_Q s^{2} - 1620 u^{2} v^{5} m_Q s^{2} - 1728 u^{2} v^{3} m_Q^{3} s \\
&\qquad - 1152 u v^{4} m_Q^{3} s - 384 u^{5} v^{2} m_Q^{3} s - 1152 u^{4} v^{3} m_Q^{3} s - 960 u^{3} v^{2} m_Q^{3} s - 1152 u^{2} v^{5} m_Q^{3} s \\
&\qquad - 540 u^{2} v^{3} m_Q s^{2} - 384 u v^{6} m_Q^{3} s + 540 u^{5} v^{3} m_Q s^{2} + 1056 u^{4} v^{2} m_Q^{3} s + 540 u^{2} v^{6} m_Q s^{2} \\
&\qquad + 288 u^{2} v^{2} m_Q^{3} s + 1152 u v^{5} m_Q^{3} s + 384 u v^{3} m_Q^{3} s + 1620 u^{4} v^{4} m_Q s^{2} + 1620 u^{3} v^{5} m_Q s^{2} \\
&\qquad + 2496 u^{3} v^{3} m_Q^{3} s + 1620 u^{3} v^{3} m_Q s^{2} + 2592 u^{2} v^{4} m_Q^{3} s + 1620 u^{2} v^{4} m_Q s^{2}\Bigr].
\end{align*}
\Needspace{6\baselineskip}
\begin{align*}
&D_7=\frac{1}{\pi^{4}}\Bigl[- 1960 u^{5} v^{4} m_Q^{3} s^{2} - 2760 u^{4} v^{5} m_Q^{3} s^{2} - 1680 u^{4} v^{5} m_Q s^{3} - 2040 u^{3} v^{6} m_Q^{3} s^{2} \\
&\qquad - 840 u^{3} v^{6} m_Q s^{3} - 1000 u^{3} v^{4} m_Q^{3} s^{2} - 600 u^{6} v^{3} m_Q^{3} s^{2} - 840 u^{5} v^{6} m_Q s^{3} \\
&\qquad - 840 u^{5} v^{4} m_Q s^{3} - 732 u^{3} v^{4} m_Q^{5} s - 640 u^{2} v^{7} m_Q^{3} s^{2} - 640 u^{2} v^{5} m_Q^{3} s^{2} - 560 u^{6} v^{5} m_Q s^{3} \\
&\qquad - 560 u^{4} v^{7} m_Q s^{3} - 360 u^{4} v^{5} m_Q^{5} s - 480 u^{4} v^{3} m_Q^{5} s - 520 u^{4} v^{3} m_Q^{3} s^{2} - 576 u^{2} v^{5} m_Q^{5} s \\
&\qquad - 144 u^{6} v^{3} m_Q^{5} s - 288 u^{5} v^{4} m_Q^{5} s - 144 u^{5} v^{2} m_Q^{5} s - 288 u^{3} v^{6} m_Q^{5} s - 144 u^{2} v^{7} m_Q^{5} s \\
&\qquad - 216 u v^{6} m_Q^{5} s - 140 u^{7} v^{4} m_Q s^{3} - 140 u^{3} v^{8} m_Q s^{3} - 140 u^{3} v^{4} m_Q s^{3} - 48 u^{2} v^{3} m_Q^{5} s \\
&\qquad - 36 u v^{8} m_Q^{5} s - 36 u v^{4} m_Q^{5} s - 36 u^{7} v^{2} m_Q^{5} s + 120 u^{6} v^{2} m_Q^{5} s + 72 u^{4} v^{2} m_Q^{5} s \\
&\qquad + 120 u^{3} v^{3} m_Q^{3} s^{2} - 12 u^{3} v^{2} m_Q^{5} s + 144 u v^{5} m_Q^{5} s + 160 u^{7} v^{3} m_Q^{3} s^{2} + 240 u^{3} v^{3} m_Q^{5} s \\
&\qquad + 160 u^{2} v^{8} m_Q^{3} s^{2} + 288 u^{2} v^{4} m_Q^{5} s + 160 u^{2} v^{4} m_Q^{3} s^{2} + 144 u v^{7} m_Q^{5} s + 560 u^{6} v^{4} m_Q s^{3} \\
&\qquad + 432 u^{5} v^{3} m_Q^{5} s + 560 u^{4} v^{4} m_Q s^{3} + 560 u^{3} v^{7} m_Q s^{3} + 560 u^{3} v^{5} m_Q s^{3} + 480 u^{2} v^{6} m_Q^{5} s \\
&\qquad + 640 u^{6} v^{4} m_Q^{3} s^{2} + 840 u^{5} v^{3} m_Q^{3} s^{2} + 768 u^{4} v^{4} m_Q^{5} s + 640 u^{3} v^{7} m_Q^{3} s^{2} \\
&\qquad + 792 u^{3} v^{5} m_Q^{5} s + 960 u^{2} v^{6} m_Q^{3} s^{2} + 1120 u^{5} v^{5} m_Q^{3} s^{2} + 1680 u^{5} v^{5} m_Q s^{3} \\
&\qquad + 1120 u^{4} v^{6} m_Q^{3} s^{2} + 1680 u^{4} v^{6} m_Q s^{3} + 2160 u^{4} v^{4} m_Q^{3} s^{2} + 2280 u^{3} v^{5} m_Q^{3} s^{2}\Bigr].
\end{align*}
\begin{equation*} E_7=E_7^{(1)}+E_7^{(2)}+E_7^{(3)}.\end{equation*}
\Needspace{6\baselineskip}
\begin{align*}
&E_7^{(1)}=\frac{1}{\pi^{4}}\Bigl[- 300 u^{6} v^{7} m_Q s^{4} - 496 u^{6} v^{5} m_Q^{3} s^{3} - 320 u^{5} v^{6} m_Q^{5} s^{2} - 720 u^{5} v^{6} m_Q^{3} s^{3} \\
&\qquad - 260 u^{4} v^{7} m_Q^{5} s^{2} - 528 u^{4} v^{7} m_Q^{3} s^{3} - 200 u^{7} v^{6} m_Q s^{4} - 224 u^{6} v^{7} m_Q^{3} s^{3} \\
&\qquad - 248 u^{6} v^{5} m_Q^{5} s^{2} - 200 u^{5} v^{8} m_Q s^{4} - 200 u^{5} v^{6} m_Q s^{4} - 210 u^{4} v^{5} m_Q^{5} s^{2} - 176 u^{7} v^{6} m_Q^{3} s^{3} \\
&\qquad - 144 u^{7} v^{4} m_Q^{3} s^{3} - 176 u^{5} v^{8} m_Q^{3} s^{3} - 132 u^{5} v^{4} m_Q^{5} s^{2} - 160 u^{3} v^{8} m_Q^{3} s^{3} \\
&\qquad - 168 u^{3} v^{6} m_Q^{5} s^{2} - 116 u^{7} v^{4} m_Q^{5} s^{2} - 100 u^{6} v^{5} m_Q s^{4} - 100 u^{4} v^{7} m_Q s^{4} - 128 u^{4} v^{5} m_Q^{3} s^{3} \\
&\qquad - 128 u^{3} v^{8} m_Q^{5} s^{2} - 80 u^{3} v^{6} m_Q^{3} s^{3} - 80 u^{8} v^{5} m_Q^{3} s^{3} - 64 u^{5} v^{4} m_Q^{3} s^{3} - 80 u^{4} v^{9} m_Q^{3} s^{3}\Bigr].
\end{align*}
\Needspace{6\baselineskip}
\begin{align*}
&E_7^{(2)}=\frac{1}{\pi^{4}}\Bigl[- 50 u^{8} v^{5} m_Q s^{4} - 26 u^{8} v^{3} m_Q^{5} s^{2} - 36 u^{6} v^{3} m_Q^{5} s^{2} - 50 u^{4} v^{9} m_Q s^{4} \\
&\qquad - 30 u^{2} v^{9} m_Q^{5} s^{2} - 60 u^{2} v^{7} m_Q^{5} s^{2} - 16 u^{9} v^{4} m_Q^{3} s^{3} - 10 u^{4} v^{5} m_Q s^{4} - 2 u^{4} v^{3} m_Q^{5} s^{2} \\
&\qquad - 16 u^{3} v^{10} m_Q^{3} s^{3} - 8 u^{3} v^{4} m_Q^{5} s^{2} - 6 u^{2} v^{5} m_Q^{5} s^{2} + 10 u^{9} v^{5} m_Q s^{4} + 6 u^{9} v^{3} m_Q^{5} s^{2} \\
&\qquad + 14 u^{5} v^{3} m_Q^{5} s^{2} + 10 u^{4} v^{10} m_Q s^{4} + 12 u^{4} v^{4} m_Q^{3} s^{3} + 6 u^{2} v^{10} m_Q^{5} s^{2} \\
&\qquad + 30 u^{8} v^{4} m_Q^{5} s^{2} + 44 u^{7} v^{3} m_Q^{5} s^{2} + 50 u^{4} v^{6} m_Q s^{4} + 30 u^{3} v^{9} m_Q^{5} s^{2} + 16 u^{3} v^{5} m_Q^{3} s^{3} \\
&\qquad + 30 u^{2} v^{6} m_Q^{5} s^{2} + 50 u^{8} v^{6} m_Q s^{4} + 50 u^{5} v^{9} m_Q s^{4} + 50 u^{5} v^{5} m_Q s^{4}\Bigr].
\end{align*}
\Needspace{6\baselineskip}
\begin{align*}
&E_7^{(3)}=\frac{1}{\pi^{4}}\Bigl[76 u^{8} v^{4} m_Q^{3} s^{3} + 72 u^{7} v^{5} m_Q^{5} s^{2} + 72 u^{4} v^{8} m_Q^{5} s^{2} + 50 u^{4} v^{4} m_Q^{5} s^{2} + 62 u^{3} v^{5} m_Q^{5} s^{2} \\
&\qquad + 60 u^{2} v^{8} m_Q^{5} s^{2} + 100 u^{7} v^{7} m_Q s^{4} + 100 u^{7} v^{5} m_Q s^{4} + 100 u^{6} v^{8} m_Q s^{4} + 108 u^{5} v^{7} m_Q^{5} s^{2} \\
&\qquad + 100 u^{4} v^{8} m_Q s^{4} + 80 u^{3} v^{9} m_Q^{3} s^{3} + 108 u^{6} v^{6} m_Q^{5} s^{2} + 176 u^{6} v^{4} m_Q^{5} s^{2} + 136 u^{6} v^{4} m_Q^{3} s^{3} \\
&\qquad + 300 u^{5} v^{7} m_Q s^{4} + 212 u^{3} v^{7} m_Q^{5} s^{2} + 160 u^{3} v^{7} m_Q^{3} s^{3} + 320 u^{7} v^{5} m_Q^{3} s^{3} \\
&\qquad + 300 u^{6} v^{6} m_Q s^{4} + 330 u^{5} v^{5} m_Q^{5} s^{2} + 368 u^{5} v^{5} m_Q^{3} s^{3} + 332 u^{4} v^{8} m_Q^{3} s^{3} \\
&\qquad + 350 u^{4} v^{6} m_Q^{5} s^{2} + 584 u^{6} v^{6} m_Q^{3} s^{3} + 592 u^{5} v^{7} m_Q^{3} s^{3} + 392 u^{4} v^{6} m_Q^{3} s^{3}\Bigr].
\end{align*}

\textbf{Heavy-line gluonic contributions.}
The reduced variables $t_3$ and $K_{3}$ are those of Eq. \eqref{eq:app-t3-k3}. The only additional heavy-line denominator polynomial is
\begin{equation}
 \Huv=u^2+uv-u+v^2-v.
 \label{eq:app-Huv}
\end{equation} The scalar strange-condensate and the $m_s\slashed x$ vector-derivative pieces are
\begin{align}
\rho_{7,\mathrm{heavy},S}^{(2)}&=\langle\bar ss\rangle\langle g_s^2G^2\rangle
\Bigl[-\frac1{15}(\mathcal T_{GG1S}+\mathcal T_{GG2S}+\mathcal T_{GG3S}) \notag\\
&\hspace{25mm}+\frac1{720}\mathcal T_{G1G2S}
-\frac1{1440}(\mathcal T_{G1G3S}+\mathcal T_{G2G3S})\Bigr],\\
\rho_{7,\mathrm{heavy},V}^{(2)}&=m_s\langle\bar ss\rangle\langle g_s^2G^2\rangle
\Bigl[\frac1{60}(\mathcal T_{GG1Vder}+\mathcal T_{GG2Vder}+\mathcal T_{GG3Vder}) \notag\\
&\hspace{25mm}-\frac1{2880}\mathcal T_{G1G2Vder}
+\frac1{5760}(\mathcal T_{G1G3Vder}+\mathcal T_{G2G3Vder})\Bigr].
\end{align}
where each $\mathcal T$ denotes an explicitly evaluated heavy-line gluonic topology. The vector derivative is obtained from the Fourier identity $x_\lambda e^{-ikx}=i\partial e^{-ikx}/\partial k^\lambda$ with the covariant contraction $\gamma^\lambda\partial/\partial k^\lambda$. The equivalent lowered-index form, $\gamma_\lambda\partial/\partial k_\lambda$, gives the same coefficients exactly.

\textbf{Scalar heavy-line topology $GG1$.}
\begin{align*}
\mathcal T_{GG1S}(s)&= \frac{40 v s u^{3} m_Q^{3} \left(u + v\right) \left(u + v - 1\right)^{2}}{3 \Huv^{2}}\,\delta^{(1)}(K_{3}) + \frac{40 u^{3} m_Q^{3} \left(u - v - 1\right)}{3 \Huv^{2}}\,\delta(K_{3}).
\end{align*}

\textbf{Scalar heavy-line topology $GG2$.}
\begin{align*}
\mathcal T_{GG2S}(s)&= \frac{40 u s v^{3} m_Q^{3} \left(u + v\right) \left(u + v - 1\right)^{2}}{3 \Huv^{2}}\,\delta^{(1)}(K_{3}) + \frac{-40 v^{3} m_Q^{3} \left(u - v + 1\right)}{3 \Huv^{2}}\,\delta(K_{3}).
\end{align*}

\textbf{Scalar heavy-line topology $GG3$.}
\begin{align*}
\mathcal T_{GG3S}(s)&= \frac{-40 \left(u + v - 1\right)^{3} m_Q s u^{2} v^{2} \mathcal Q_{GG3S,1}}{3 \Huv^{4}}\,\delta^{(1)}(K_{3}) + \frac{40 \left(u + v - 1\right)^{3} m_Q \mathcal Q_{GG3S,2}}{3 \Huv^{4}}\,\delta(K_{3}) \\
&\quad + \frac{-40 m_Q \left(u + v - 1\right)^{3} \mathcal Q_{GG3S,3}}{3 \Huv^{4}}\,\Theta(K_{3}).
\end{align*}

\textbf{Scalar heavy-line topology $G1G2$.}
\begin{equation*}
\mathcal T_{G1G2S}(s)= \frac{320 m_Q s u^{2} v^{2} \left(u + v - 1\right)^{2}}{\Huv^{3}}\,\delta(K_{3})+\frac{-960 u v m_Q \left(u + v - 1\right)}{\Huv^{3}}\,\Theta(K_{3}).
\end{equation*}

\textbf{Scalar heavy-line topology $G1G3$.}
\begin{align*}
\mathcal T_{G1G3S}(s)&= \frac{320 \left(u + v - 1\right) u m_Q \mathcal Q_{G1G3S,4}}{\Huv^{3}}\,\delta(K_{3}) + \frac{-320 u m_Q \left(u + v - 1\right)^{2}}{\Huv^{3}}\,\Theta(K_{3}).
\end{align*}

\textbf{Scalar heavy-line topology $G2G3$.}
\begin{align*}
\mathcal T_{G2G3S}(s)&= \frac{320 \left(u + v - 1\right) v m_Q \mathcal Q_{G2G3S,5}}{\Huv^{3}}\,\delta(K_{3}) + \frac{-320 v m_Q \left(u + v - 1\right)^{2}}{\Huv^{3}}\,\Theta(K_{3}).
\end{align*}

\textbf{Vector-derivative heavy-line topology $GG1$.}
\begin{align*}
\mathcal T_{GG1Vder}(s)&= \frac{80 v s u^{3} m_Q^{4} \left(u + v\right) \left(u + v - 1\right)^{3}}{3 \Huv}\,\delta^{(2)}(K_{3}) \\
&\quad + \frac{80 \left(u + v - 1\right) m_Q^{2} u^{3} \mathcal Q_{GG1Vder,6}}{3 \Huv^{3}}\,\delta^{(1)}(K_{3}) \\
&\quad + \frac{80 m_Q^{2} u^{3} \left(u + v - 1\right) \mathcal Q_{GG1Vder,7}}{3 \Huv^{3}}\,\delta(K_{3}).
\end{align*}

\textbf{Vector-derivative heavy-line topology $GG2$.}
\begin{align*}
\mathcal T_{GG2Vder}(s)&= \frac{80 u s v^{3} m_Q^{4} \left(u + v\right) \left(u + v - 1\right)^{3}}{3 \Huv}\,\delta^{(2)}(K_{3}) \\
&\quad + \frac{80 \left(u + v - 1\right) m_Q^{2} v^{3} \mathcal Q_{GG2Vder,8}}{3 \Huv^{3}}\,\delta^{(1)}(K_{3}) \\
&\quad - \frac{80 m_Q^{2} v^{3} \left(u + v - 1\right) \mathcal Q_{GG2Vder,9}}{3 \Huv^{3}}\,\delta(K_{3}).
\end{align*}

\textbf{Vector-derivative heavy-line topology $GG3$.}
\begin{align*}
\mathcal T_{GG3Vder}(s)&= -\frac{20 \left(u + v - 1\right)^{4} m_Q^{2} \mathcal Q_{GG3Vder,10} \mathcal Q_{GG3Vder,11}}{3 \Huv^{3}}\,\delta^{(2)}(K_{3}) \\
&\quad + \frac{20 \left(u + v - 1\right)^{4} m_Q^{2} \mathcal Q_{GG3Vder,12}}{3 \Huv^{3}}\,\delta^{(1)}(K_{3}) \\
&\quad - \frac{20 m_Q^{2} \left(u + v - 1\right)^{4} \mathcal Q_{GG3Vder,13}}{3 \Huv^{3}}\,\delta(K_{3}).
\end{align*}

\textbf{Vector-derivative heavy-line topology $G1G2$.}
\begin{align*}
\mathcal T_{G1G2Vder}(s)&= \frac{640 s u^{2} v^{2} m_Q^{2} \left(u + v - 1\right)^{3}}{\Huv^{2}}\,\delta^{(1)}(K_{3}) + \frac{640 \left(u + v - 1\right)^{2} u v \mathcal Q_{G1G2Vder,14}}{\Huv^{4}}\,\delta(K_{3}) \\
&\quad + \frac{-640 u v \left(u + v - 1\right) \mathcal Q_{G1G2Vder,15}}{\Huv^{4}}\,\Theta(K_{3}).
\end{align*}

\textbf{Vector-derivative heavy-line topology $G1G3$.}
\begin{align*}
\mathcal T_{G1G3Vder}(s)&= \frac{320 \left(u + v - 1\right)^{2} u \mathcal Q_{G1G3Vder,16} \mathcal Q_{G1G3Vder,17}}{\Huv^{4}}\,\delta^{(1)}(K_{3}) \\
&\quad + \frac{-320 \left(u + v - 1\right)^{3} u \mathcal Q_{G1G3Vder,18}}{\Huv^{4}}\,\delta(K_{3}) + \frac{960 v u^{2} \left(u + v - 1\right)^{2}}{\Huv^{4}}\,\Theta(K_{3}).
\end{align*}

\textbf{Vector-derivative heavy-line topology $G2G3$.}
\begin{align*}
\mathcal T_{G2G3Vder}(s)&= \frac{320 \left(u + v - 1\right)^{2} v \mathcal Q_{G2G3Vder,19} \mathcal Q_{G2G3Vder,20}}{\Huv^{4}}\,\delta^{(1)}(K_{3}) \\
&\quad + \frac{-320 \left(u + v - 1\right)^{3} v \mathcal Q_{G2G3Vder,21}}{\Huv^{4}}\,\delta(K_{3}) + \frac{960 u v^{2} \left(u + v - 1\right)^{2}}{\Huv^{4}}\,\Theta(K_{3}).
\end{align*}

Expanded auxiliary polynomials appearing above.
\begin{align*}
\mathcal Q_{GG3S,1} &= u^{2} m_Q^{2} + u^{4} m_Q^{2} + v^{2} m_Q^{2} + v^{4} m_Q^{2} - 2 u^{3} m_Q^{2} - 2 v^{3} m_Q^{2} + u v s + u v^{3} s + u^{3} v s - 4 u v^{2} m_Q^{2} \\
&\quad - 4 u^{2} v m_Q^{2} - 2 u v^{2} s - 2 u^{2} v s + 2 u v m_Q^{2} + 2 u v^{3} m_Q^{2} + 2 u^{3} v m_Q^{2} + 2 u^{2} v^{2} s + 3 u^{2} v^{2} m_Q^{2}
\end{align*}
\begin{align*}
\mathcal Q_{GG3S,2} &= - 4 u^{4} m_Q^{2} - 4 v^{4} m_Q^{2} + 2 u^{3} m_Q^{2} + 2 u^{5} m_Q^{2} + 2 v^{3} m_Q^{2} + 2 v^{5} m_Q^{2} - 16 u^{2} v^{2} m_Q^{2} - 12 u v^{3} m_Q^{2} \\
&\quad - 12 u^{3} v m_Q^{2} - 11 u^{2} v^{2} s - 4 u v^{3} s - 4 u^{3} v s + 2 u v^{2} s + 2 u v^{4} s + 2 u^{2} v s + 2 u^{4} v s + 6 u v^{2} m_Q^{2} \\
&\quad + 6 u v^{4} m_Q^{2} + 6 u^{2} v m_Q^{2} + 6 u^{4} v m_Q^{2} + 9 u^{2} v^{3} s + 9 u^{3} v^{2} s + 10 u^{2} v^{3} m_Q^{2} + 10 u^{3} v^{2} m_Q^{2}
\end{align*}
\begin{align*}
\mathcal Q_{GG3S,3} &= - 2 u - 2 v + 2 u^{2} + 2 v^{2} + 5 u v
\end{align*}
\begin{align*}
\mathcal Q_{G1G3S,4} &= u^{2} m_Q^{2} + u^{4} m_Q^{2} + v^{2} m_Q^{2} + v^{4} m_Q^{2} - 2 u^{3} m_Q^{2} - 2 v^{3} m_Q^{2} + u v s + u v^{3} s + u^{3} v s - 4 u v^{2} m_Q^{2} \\
&\quad - 4 u^{2} v m_Q^{2} - 2 u v^{2} s - 2 u^{2} v s + 2 u v m_Q^{2} + 2 u v^{3} m_Q^{2} + 2 u^{3} v m_Q^{2} + 2 u^{2} v^{2} s + 3 u^{2} v^{2} m_Q^{2}
\end{align*}
\begin{align*}
\mathcal Q_{G2G3S,5} &= u^{2} m_Q^{2} + u^{4} m_Q^{2} + v^{2} m_Q^{2} + v^{4} m_Q^{2} - 2 u^{3} m_Q^{2} - 2 v^{3} m_Q^{2} + u v s + u v^{3} s + u^{3} v s - 4 u v^{2} m_Q^{2} \\
&\quad - 4 u^{2} v m_Q^{2} - 2 u v^{2} s - 2 u^{2} v s + 2 u v m_Q^{2} + 2 u v^{3} m_Q^{2} + 2 u^{3} v m_Q^{2} + 2 u^{2} v^{2} s + 3 u^{2} v^{2} m_Q^{2}
\end{align*}
\begin{align*}
\mathcal Q_{GG1Vder,6} &= v^{3} s + v^{5} s + u^{5} m_Q^{2} + v^{3} m_Q^{2} + v^{4} m_Q^{2} - u^{2} m_Q^{2} - v^{2} m_Q^{2} - v^{5} m_Q^{2} \\
&\quad - 3 u^{4} m_Q^{2} - 2 v^{4} s + 3 u^{3} m_Q^{2} + u^{2} v s + u^{4} v s + u^{4} v m_Q^{2} + u^{3} v^{2} m_Q^{2} \\
&\quad - u v^{4} m_Q^{2} - u^{2} v^{3} m_Q^{2} - 6 u v^{3} s - 6 u^{2} v^{2} s - 4 u^{3} v m_Q^{2} - 3 u^{2} v^{2} m_Q^{2} - 2 u v m_Q^{2} \\
&\quad - 2 u^{3} v s + 2 u v^{2} s + 3 u v^{2} m_Q^{2} + 4 u v^{4} s + 4 u^{3} v^{2} s + 5 u^{2} v m_Q^{2} + 6 u^{2} v^{3} s
\end{align*}
\begin{align*}
\mathcal Q_{GG1Vder,7} &= u^{2} - u - v - 2 v^{2}
\end{align*}
\begin{align*}
\mathcal Q_{GG2Vder,8} &= u^{3} s + u^{5} s + u^{3} m_Q^{2} + u^{4} m_Q^{2} + v^{5} m_Q^{2} - u^{2} m_Q^{2} - u^{5} m_Q^{2} \\
&\quad - v^{2} m_Q^{2} - 3 v^{4} m_Q^{2} - 2 u^{4} s + 3 v^{3} m_Q^{2} + u v^{2} s + u v^{4} s + u v^{4} m_Q^{2} + u^{2} v^{3} m_Q^{2} \\
&\quad - u^{4} v m_Q^{2} - u^{3} v^{2} m_Q^{2} - 6 u^{3} v s - 6 u^{2} v^{2} s - 4 u v^{3} m_Q^{2} - 3 u^{2} v^{2} m_Q^{2} - 2 u v m_Q^{2} \\
&\quad - 2 u v^{3} s + 2 u^{2} v s + 3 u^{2} v m_Q^{2} + 4 u^{4} v s + 4 u^{2} v^{3} s + 5 u v^{2} m_Q^{2} + 6 u^{3} v^{2} s
\end{align*}
\begin{align*}
\mathcal Q_{GG2Vder,9} &= u + v - v^{2} + 2 u^{2}
\end{align*}
\begin{align*}
\mathcal Q_{GG3Vder,10} &= u^{2} m_Q^{2} + u^{4} m_Q^{2} + v^{2} m_Q^{2} + v^{4} m_Q^{2} - 2 u^{3} m_Q^{2} - 2 v^{3} m_Q^{2} + u v s + u v^{3} s + u^{3} v s - 4 u v^{2} m_Q^{2} \\
&\quad - 4 u^{2} v m_Q^{2} - 2 u v^{2} s - 2 u^{2} v s + 2 u v m_Q^{2} + 2 u v^{3} m_Q^{2} + 2 u^{3} v m_Q^{2} + 2 u^{2} v^{2} s + 3 u^{2} v^{2} m_Q^{2}
\end{align*}
\begin{align*}
\mathcal Q_{GG3Vder,11} &= u^{2} m_Q^{2} + u^{4} m_Q^{2} + v^{2} m_Q^{2} + v^{4} m_Q^{2} - 2 u^{3} m_Q^{2} - 2 v^{3} m_Q^{2} - 4 u v^{2} m_Q^{2} \\
&\quad - 4 u^{2} v m_Q^{2} + 2 u v m_Q^{2} + 2 u v^{3} m_Q^{2} + 2 u^{3} v m_Q^{2} + 3 u^{2} v^{2} s + 3 u^{2} v^{2} m_Q^{2}
\end{align*}
\begin{align*}
\mathcal Q_{GG3Vder,12} &= - u^{2} m_Q^{2} - v^{2} m_Q^{2} - 7 u^{4} m_Q^{2} - 7 v^{4} m_Q^{2} + 3 u^{5} m_Q^{2} + 3 v^{5} m_Q^{2} + 5 u^{3} m_Q^{2} \\
&\quad + 5 v^{3} m_Q^{2} - 27 u^{2} v^{2} m_Q^{2} - 20 u v^{3} m_Q^{2} - 20 u^{3} v m_Q^{2} - 17 u^{2} v^{2} s - 4 u v^{3} s \\
&\quad - 4 u^{3} v s - 2 u v m_Q^{2} + 2 u v^{2} s + 2 u v^{4} s + 2 u^{2} v s + 2 u^{4} v s + 9 u v^{4} m_Q^{2} + 9 u^{4} v m_Q^{2} \\
&\quad + 13 u v^{2} m_Q^{2} + 13 u^{2} v m_Q^{2} + 15 u^{2} v^{3} s + 15 u^{3} v^{2} s + 15 u^{2} v^{3} m_Q^{2} + 15 u^{3} v^{2} m_Q^{2}
\end{align*}
\begin{align*}
\mathcal Q_{GG3Vder,13} &= - 2 u - 2 v + 2 u^{2} + 2 v^{2} + 7 u v
\end{align*}
\begin{align*}
\mathcal Q_{G1G2Vder,14} &= - 3 u^{2} m_Q^{2} - 3 u^{4} m_Q^{2} - 3 v^{2} m_Q^{2} - 3 v^{4} m_Q^{2} + 6 u^{3} m_Q^{2} \\
&\quad + 6 v^{3} m_Q^{2} + u v^{3} s + u^{3} v s - u v^{2} s - u^{2} v s - 9 u^{2} v^{2} m_Q^{2} - 6 u v m_Q^{2} \\
&\quad - 6 u v^{3} m_Q^{2} - 6 u^{3} v m_Q^{2} + 2 u^{2} v^{2} s + 12 u v^{2} m_Q^{2} + 12 u^{2} v m_Q^{2}
\end{align*}
\begin{align*}
\mathcal Q_{G1G2Vder,15} &= - 3 u - 3 v + 3 u^{2} + 3 v^{2} + 8 u v
\end{align*}
\begin{align*}
\mathcal Q_{G1G3Vder,16} &= u^{2} m_Q^{2} + u^{4} m_Q^{2} + v^{2} m_Q^{2} + v^{4} m_Q^{2} - 2 u^{3} m_Q^{2} - 2 v^{3} m_Q^{2} \\
&\quad + u v s + u v^{3} s + u^{3} v s - 4 u v^{2} m_Q^{2} - 4 u^{2} v m_Q^{2} - 2 u v^{2} s - 2 u^{2} v s \\
&\quad + 2 u v m_Q^{2} + 2 u v^{3} m_Q^{2} + 2 u^{3} v m_Q^{2} + 2 u^{2} v^{2} s + 3 u^{2} v^{2} m_Q^{2}
\end{align*}
\begin{align*}
\mathcal Q_{G1G3Vder,17} &= u^{2} m_Q^{2} + u^{4} m_Q^{2} + v^{2} m_Q^{2} + v^{4} m_Q^{2} - 2 u^{3} m_Q^{2} - 2 v^{3} m_Q^{2} + u^{2} v^{2} s \\
&\quad - 4 u v^{2} m_Q^{2} - 4 u^{2} v m_Q^{2} + 2 u v m_Q^{2} + 2 u v^{3} m_Q^{2} + 2 u^{3} v m_Q^{2} + 3 u^{2} v^{2} m_Q^{2}
\end{align*}
\begin{align*}
\mathcal Q_{G1G3Vder,18} &= u^{2} m_Q^{2} + u^{4} m_Q^{2} + v^{2} m_Q^{2} + v^{4} m_Q^{2} - 2 u^{3} m_Q^{2} - 2 v^{3} m_Q^{2} - 4 u v^{2} m_Q^{2} \\
&\quad - 4 u^{2} v m_Q^{2} + 2 u v m_Q^{2} + 2 u v^{3} m_Q^{2} + 2 u^{3} v m_Q^{2} + 3 u^{2} v^{2} m_Q^{2} + 5 u^{2} v^{2} s
\end{align*}
\begin{align*}
\mathcal Q_{G2G3Vder,19} &= u^{2} m_Q^{2} + u^{4} m_Q^{2} + v^{2} m_Q^{2} + v^{4} m_Q^{2} - 2 u^{3} m_Q^{2} - 2 v^{3} m_Q^{2} \\
&\quad + u v s + u v^{3} s + u^{3} v s - 4 u v^{2} m_Q^{2} - 4 u^{2} v m_Q^{2} - 2 u v^{2} s - 2 u^{2} v s \\
&\quad + 2 u v m_Q^{2} + 2 u v^{3} m_Q^{2} + 2 u^{3} v m_Q^{2} + 2 u^{2} v^{2} s + 3 u^{2} v^{2} m_Q^{2}
\end{align*}
\begin{align*}
\mathcal Q_{G2G3Vder,20} &= u^{2} m_Q^{2} + u^{4} m_Q^{2} + v^{2} m_Q^{2} + v^{4} m_Q^{2} - 2 u^{3} m_Q^{2} - 2 v^{3} m_Q^{2} + u^{2} v^{2} s \\
&\quad - 4 u v^{2} m_Q^{2} - 4 u^{2} v m_Q^{2} + 2 u v m_Q^{2} + 2 u v^{3} m_Q^{2} + 2 u^{3} v m_Q^{2} + 3 u^{2} v^{2} m_Q^{2}
\end{align*}
\begin{align*}
\mathcal Q_{G2G3Vder,21} &= u^{2} m_Q^{2} + u^{4} m_Q^{2} + v^{2} m_Q^{2} + v^{4} m_Q^{2} - 2 u^{3} m_Q^{2} - 2 v^{3} m_Q^{2} - 4 u v^{2} m_Q^{2} \\
&\quad - 4 u^{2} v m_Q^{2} + 2 u v m_Q^{2} + 2 u v^{3} m_Q^{2} + 2 u^{3} v m_Q^{2} + 3 u^{2} v^{2} m_Q^{2} + 5 u^{2} v^{2} s
\end{align*}

Distributional conversion in the present notation. For comparison with alternative Appendix conventions such as Ref. \cite{Xu:2025}, derivatives of $\delta(K_{3})$ may be eliminated only after the Borel transformation and the $s$ integration:
\begin{align}
F\delta^{(1)}(K_{3})&\doteq\frac1{p_3}\left(\frac{F}{M_B^2}-\partial_sF\right)\delta(K_{3}),\\
F\delta^{(2)}(K_{3})&\doteq\frac1{p_3^2}\left(\partial_s^2F-\frac{2}{M_B^2}\partial_sF+\frac{F}{M_B^4}\right)\delta(K_{3}),\\
F\delta^{(3)}(K_{3})&\doteq\frac1{p_3^3}\left(-\partial_s^3F+\frac{3}{M_B^2}\partial_s^2F-\frac{3}{M_B^4}\partial_sF+\frac{F}{M_B^6}\right)\delta(K_{3}).
\end{align}
where $p_3=uv\,t_3$ and $\doteq$ denotes equality after integration with $e^{-s/M_B^2}$.  Thus the strict spin-2 distributional formulas above can be mapped term by term to other post-Borel conventions without changing the underlying spectral density.

% ===========================================================================

\bibliographystyle{unsrt}
\bibliography{references}

\end{document}